\documentstyle[preprint,prl,aps]{revtex}
\begin{document}
\draft

\preprint{\vbox{\hbox{July 1996}\hbox{IFP-729-UNC}}}

\title{The Three Families from $SU(4)_A\otimes SU(3)_C\otimes SU(2)_L\otimes U(1)_X$ SM-like Chiral Models.}
\author{\bf  Otto C. W. Kong}
\address{Institute of Field Physics, Department of Physics and Astronomy,\\
University of North Carolina, Chapel Hill, NC  27599-3255}
\maketitle

\begin{abstract}
We give a detailed description of the model construction procedures about our
new approach to the family structure of the standard model.  SM-like chiral 
fermion spectra, largely "derivable" from the gauge anomaly constraints, are
formulated in a   $SU(N)\otimes SU(3)\otimes SU(2)\otimes U(1)$ symmetry 
framework as an extension of the SM symmetry. The $N=4$ case gives naturally
three families as a result, with $U(1)_Y$ nontrivially embedded into the 
$SU(4)_A\otimes U(1)_X$. Such a spectrum has extra vector-like quarks and
leptons. We illustrate how an acceptable symmetry breaking pattern can be obtained
through a relatively simple scalar sector which gives naturally hierarchical
quark mass matrices. Compatibility with various FCNC constraints and some
interesting aspects of the possible phenomenological features are discussed,
from a non-model specific perspective. The question of incorporating supersymmetry
without putting in the Higgses as extra supermultiplet is also addressed.

\end{abstract}


\newpage

\section{INTRODUCTION}

The matter content of the phenomenologically very successful
standard model(SM) consists of three sets (families) of 15 chiral fermion
states of identical quantum numbers. In addition, a scalar doublet
Higgs is needed to break the electroweak(EW) symmetry and give the fermions
masses. The existence of the scalar leads
to the hierarchy problem, which is widely believed to be addressed by
supersymmetry(SUSY). SUSY then doubles the particle spectrum, giving,
in particular, scalar partners to fermions. Moreover, one extra Higgs
doublet is needed; and the Higgses get fermionic partners. This makes up the
minimal supersymmetric SM (MSSM). The representation structure
of a single family of chiral fermions  is very strongly constrained by
the requirement of cancellation of all  gauge anomalies, making it
easily "derivable" once some simple assumptions are taken\cite{sm1,sm2}.
However, the existence of {\it three} families, with the great hierarchy of
masses among the fermions after EW-symmetry breaking, remains a
mystery.

In a short letter earlier\cite{728}, we introduced a new approach to this
mysterious family problem. The approach is based on an attempt to
mimic the highly contrained group representation structure of the
one-family SM fermion spectrum while extending the gauge symmetry.
The motivation is to look for a similarly structured list of chiral
representations that, upon breaking of the extra symmetry,
gives the three SM families naturally as the low-energy chiral spectrum.
A general $SU(N)\otimes SU(3)\otimes SU(2)\otimes U(1)$ gauge symmetry was
considered, with a successful
$SU(4)_A\otimes SU(3)_C\otimes SU(2)_L\otimes U(1)_X$ model presented. A comparison
with  other approaches to the family problem was also given. A more
ambitious idea also discussed is to incorporate SUSY without
introducing extra chiral supermultiplets.

Here in this paper, we would like first to report on the details of
our model construction procedures (section-II). The fact that the two essential
desirable features, namely a chiral spectrum similarly structured
as the one-family SM and free from all  anomalies and a
natural embedding of the three SM families, can be incorporated into
one model is very nontrivial. However, our approach does have some flexibility that allows modifications on a basic framework. We
will  present  four explicit 
$SU(4)_A\otimes SU(3)_C\otimes SU(2)_L\otimes U(1)_X$ models obtained from
slightly different modifications. Some comments on their relative merits
can be found in the subsequent sections.

The models hold promises for an interesting phenomenology, provided
that at least some of the extra symmetries are broken at a
relative low energy scale. We will also try here to take a first
look into the possible phenomenological features. However, we will try to
confine our discussions mainly to the more general features of the approach, instead 
of the specifically model-dependent ones. Our aim is to illustrate how this
new approach to the family structure can lead to interesting  and
potentially very successful models, and to provide useful guidelines 
for future detailed model constructions.

While our approach is essentially different from simply extending the SM
gauge symmetry with a horizontal symmetry\cite{sun,su2,su3},  part
of the extra gauge symmetry obtained do behave like horizontal symmetries
and can be analyzed from a similar phenomenological perspective.

In  section-III, we will discuss how an acceptable symmetry breaking pattern
can be obtained through a relatively simple scalar sector. We will see that
this can be naturally arranged in such a way that gives SM quark mass 
matrices with a desirable hierarchical pattern, while evading the 
flavor changing neutral current(FCNC) constraints. In section-IV, we will
look at  the gauge sector, with its FCNC constraints and some features of 
the renormalization-group(RG)-runnings of the gauge couplings; the question of incorporating SUSY; and some interesting
aspects of the leptonic sector.  We will then make some concluding remarks in
the last section.

\section{SM-LIKE CHIRAL SPECTRUM}
\subsection{One family standard model}
We start with the following perspective on
the elegance of the SM representation
structure for a single family.
\begin{itemize}
\item
One can start by introducing
the simplest multiplet that transforms nontrivially under each of
the component group factors,
namely a ${\bf (3,2,1)}$ of  
$SU(3)_C\otimes SU(2)_L\otimes U(1)_Y$.(Fixing the hypercharge as 1
just corresponds to a arbitrary choice of normalization, here differs from
the standard normalization by a factor of 6. We are going to stick with this
particular normalization throughout the paper.)
\item
To cancel the
$SU(3)$ anomaly, two ${\bf \bar{3}}$'s are needed. Keeping with the chiral structure,
we have to use a ${\bf (\bar{3},1,x)}$ and a ${\bf (\bar{3},1,y)}$, with the hypercharges yet to be specified.
\item
Next, cancellation of the global-$SU(2)$ anomaly dictates the inclusion of
an extra doublet,  ${\bf (1,2,a)}$.
\item
We still have to cancel all the $U(1)$-anomalies.
We have $a=-3$ from the $[SU(2)]^2U(1)$ anomaly constraint.
$x$ and $y$ then have to satisfy  the three remaining contraints and no solution can be
found. If we then allow one singlet state, ${\bf (1,1,k)}$, we have three equations
for three unknowns.
\begin{equation}
x + y  = -2
\end{equation}
\begin{equation}
3x + 3y + k = 0
\end{equation}
\begin{equation}
3x^3 + 3y^3 + k^3 = 45
\end{equation}
This gives a unique solution, the SM hypercharge
assignment $(x, y = 2, -4; k=6)$. Notice that the solution {\it a priori} may not give a
set of rational numbers. The triviality of solution here is a bit
deceiving.
\item
A  interesting fact is: if we allow two singlets, there is a one
parameter set of solution with $x, y = 2-n, -4+n$ and  hypercharges for
the singlets given by $6-n$ and $n$ ($n$ being any integer).
\end{itemize}

\subsection{$SU(N)\otimes SU(3)\otimes SU(2)\otimes U(1)$ extensions}
We want to try to mimic the above feature in a extended symmetry that
can incorporate the three families naturally.
Consider adding one more component group factor, a
$SU(N)\otimes SU(3)\otimes SU(2)\otimes U(1)$ guage group.
$N=4$ is obviously very suggestive. Nevertheless, let us first
not fix $N$ and neglect for the moment the $U(1)$-charges. We
will also suspend about the SM-embedding, with the hope to learn more
about the general features to the kind of representation
structures.
\begin{itemize}
\item
We start with a ${\bf (N,3,2)}$ and first take the list:
${\bf (N,3,2)}, {\bf (\bar{N},\bar{3},1)}, {\bf (\bar{N},1,2)},
{\bf (\bar{N},1,1)}.$
This gives an equal number of ${\bf N}$'s and
${\bf \bar{N}}$'s and hence free us from the $SU(N)$-anomaly.
\item
The  $SU(3)$-anomaly then suggests that there is a deficiency in
$N$ ${\bf \bar{3}}$'s. If we take only one ${\bf (1,\bar{3},2)}$,
as suggested most naturally by the pattern, we would have to take
$N-2$ ${\bf (1,\bar{3},1)}$'s. In general, if we can allow $m$
 ${\bf (1,\bar{3},2)}$'s with $N-2m$ ${\bf (1,\bar{3},1)}$'s, for
any $2m\leq N$.
\item
With a fixed choice of the numbers of ${\bf (1,\bar{3},2)}$'s
and ${\bf (1,\bar{3},1)}$'s, we have just to count the number of
$SU(2)$ doublets and add one ${\bf (1,1,2)}$ if the count is odd.
That caters for the global $SU(2)$-anomaly. From this perspective,
one can always add in any extra even number of ${\bf (1,1,2)}$'s
without upsetting the anomaly constraints. The more appealing idea
is of course to stay with the a minimal content.
\item
To complete the list of representation structures, $U(1)$-charges
have to be assigned to each representations in such a way that all
the gauge anomalies involving the $U(1)$ be satisfied. However, we
would also like to add in pure singlets, ${\bf (1,1,1)}$'s.
Sticking to the SM pattern suggests taking just one, while
a model with any number of singlet states is admissible.
\item
For example, if we take $N=4$, we would arrive at the natural list:
\begin{center}
${\bf (4,3,2,1)}, {\bf (\bar{4},\bar{3},1,x)}, {\bf (\bar{4},1,2,y)},
{\bf (\bar{4},1,1,z)},$  \\
${\bf (1,\bar{3},2,a)}, {\bf (1,\bar{3},1,b)}, {\bf (1,\bar{3},1,c)},$ \\
${\bf (1,1,2,k)}, {\bf (1,1,1,s)}$.
\end{center}
The $U(1)$-related anomaly constraints give the conditions:
\begin{equation}
(3x + 2y + z) + 6 = 0 ;
\end{equation}
\begin{equation}
4x + (2a + b + c) + 8 = 0 ;
\end{equation}
\begin{equation}
4y + 3a + k + 12 =0 ;
\end{equation}
\begin{equation}
(12x + 8y + 4z) + (6a + 3b + 3c)  + 2k + s + 24 =0 ;
\end{equation}
\begin{equation}
(12x^3 + 8y^3 + 4z^3) + (6a^3 + 3b^3 + 3c^3) + 2k^3 + s^3 + 24 =0 .
\end{equation}
This is a list of five equations with eight unknowns. It looks like
there are many solutions; but again getting a set of, reasonably
small, rational numbers as a solution may be nontrivial. Simple
solution does exist, for example $(x=-2, y=-1, z=2, a=-2, b=2, c=2,
w=-2, k=4)$ satisfies the equations. However, unlike the SM case,
there is no reason to think that the solution is in any sense
unique. Obviously, situations for other $N$ values are similar.
\end{itemize}

The analysis suggests an interesting pattern of SM-like chiral spectrum similarly constrained by, or "derivable"  from the anomaly cancellation conditions. Of course we still have to build a connection to SM phenomenology.

\subsection{Embedding the three-family SM}
Here, we come back to the domain of real world particle physics. We
know that if any  $SU(N)\otimes SU(3)\otimes SU(2)\otimes U(1)$
model has to describe nature, it has to incorporate the three-family SM as a low-energy effective field theory.
\begin{itemize}
\item
A trivial embedding of the SM symmetry gives the $SU(N)$ component as
a pure horizontal symmetry. The latter is a familiar idea\cite{sun,su2,su3}, however,
a horizontal symmetry model with the kind 
of SM-like chiral spectrum from our approach has not been proposed before.
We note that  $N=3$ does give a natural three family pattern, as shown in
Table 1. However, the $[SU(3)_H]^2U(1)_Y$-anomaly cannot be canceled, 
without modifying the spectrum.
\item
The next alternative is to have only the $U(1)_Y$, the
(EW-)hypercharge, embedded nontrivially. We denote the symmetry by
$SU(N)_A\otimes SU(3)_C\otimes SU(2)_L\otimes U(1)_X$. Note that
the  $SU(N)_A$, together with the $U(1)_X$, contains the hypercharge
and is therefore {\it not} a purely horizontal symmetry. Following the argument  of
section-II.B, we see that the possible net numbers of quark doublets
and singlets  are given by $N-m$ and $2N-2m$ respectively. Once
the three families of quarks are successfully embedded, the correct
number of chiral leptons  would be easy to obtain, though  one may
have to relax the number of ${\bf (1,1,2)}$ and
${\bf (1,1,1)}$ states.
Hence, we require
\[ N-m = 3; \ \ (2N-2m = 6); \ \ 2m\leq N. \]
Acceptable solutions are given by
\[(N,m) = (4,1), \hspace{.2in} (5,2), \hspace{.2in} and \hspace{.2in} (6,3). \]
There is also an $(N,m)=(3,0)$ trivial embedding solution mentioned
above. The three solutions otherwise cannot allow a trivial
embedding. Relaxing the $2m\leq N$ criteria, and allowing
${\bf (1,3,1)}$'s instead of ${\bf (1,\bar{3},1)}$'s, then solution
is possible for all $N$. Obviously, the $N=4$ case is singled out
as the most interesting and gives three quark families most
naturally. For instance, $(N,m)=(3,1)$ gives two quark families, while
$(N,m)=(5,1)$ gives four. $N=4$ is
also the only one that gives naturally three leptonic
doublets. See Table 1. for an illustration.
\item
One can also consider nontrivial embedding of the  $SU(3)_C$ and
$SU(2)_L$ component factors. At least for $N\leq 6$, a SM-like
chiral spectrum fails to give three families.
\item
To complete the SM embedding, we need to fix the explicit
hypercharge embedding. This is done for the
$N=4$ case in the next section.
\end{itemize}

\subsection{Illustrative $SU(4)_A\otimes SU(3)_C\otimes SU(2)_L\otimes U(1)_X$ models}
Now, we stick to the most interesting case of $N=4$ and analyse all
possible embedding of $U(1)_Y$ into $SU(4)_A\otimes U(1)_X$.
\begin{itemize}
\item
There are three independent $U(1)$ subgroups in a $SU(4)$. We choose
to consider $SU(2)_H\otimes SU(2)_{K}\otimes U(1)_Z \subset SU(4)_A$
with the two extra $U(1)$ interpreted as the diagonal generators of
the two $SU(2)$'s. The $U(1)_Y$ can be any linear combination of the
three and the  $U(1)_X$. To fix a convention, we take 
${\bf 4 \longrightarrow (2,1)_{-1} + (1,2)_1}$.
\item
In Table 2, we listed the the $U(1)_Y$ states, following the list
of representations and notations of section-II.B, with
\begin{equation}
U(1)_Y = \alpha U(1)_X + \beta U(1)_Z + \gamma U(1)_H + \delta U(1)_K .
\end{equation}
The $U(1)_H$ and $U(1)_K$ here are the $U(1)$-subgroups of the correspondent $SU(2)$'s;
in particular,
\begin{equation}
U(1)_{H,K} = 2 T_{3(H,K)} \ .
\end{equation}  
To get a three-family structure for the quarks, we
can set, without lose of generosity,
\begin{equation}
\gamma = 0 ,
\end{equation}
\begin{equation}
\delta = -2\beta ,
\end{equation}
\begin{equation}
a\alpha = -\alpha - 3\beta  ,
\end{equation} and
\begin{equation}
b\alpha = c\alpha = x\alpha - 3\beta .
\end{equation}
Similarly, we obtain for the leptonic sector
\begin{equation}
k\alpha =  -y\alpha + 3\beta
\end{equation}
and
\begin{equation}
s\alpha =  -z\alpha + 3\beta .
\end{equation}
The three-family structure is shown in the last column of the table.
Note that the structure is compatible with a
\begin{center}
$SU(4)_A\otimes U(1)_X \longrightarrow SU(3)_H\otimes U(1)_{Z^{'}}\otimes U(1)_X$ \\
  \hspace{.3in} $\longrightarrow SU(3)_H\otimes U(1)_Y$
\end{center}
symmetry embedding as used in our earlier presentation\cite{728}. Both the $SU(3)_H$
or the $SU(2)_H$ may then serve as a real horizontal symmetry for
the SM under this framework.
\item
To get the correct hypercharges, we have, for the three quark
doublets
\begin{equation}
\alpha -\beta = 1 ,
\end{equation}
and for the quark singlets
\begin{equation}
x\alpha + \beta = 2 \hspace{.2in} (or \hspace{.1in} -4) ,
\end{equation}
and
\begin{equation}
x\alpha - 3\beta = -4 \hspace{.2in} (or \hspace{.1in} 2) .
\end{equation}
We denote hereafter the two different quark-singlets embeddings, given through
the two hypercharge identifications as shown in the equations, 
as schemes $I$ and {\it II} respectively.
For the leptons, we require
\begin{equation}
y\alpha + \beta = -3
\end{equation}
and
\begin{equation}
z\alpha + \beta = 6 .
\end{equation}
Each of the two schemes gives a unique solution:
\begin{eqnarray}
 I: \alpha=5/2, \beta=3/2; x=1/5, y=-9/5, z=9/5;  \nonumber  \\
 \hspace{.5in} a=-14/5, b=c=-8/5; k=18/5, s=0; \nonumber
\end{eqnarray}
\begin{eqnarray}
 II: \alpha=-1/2, \beta=-3/2; x=5, y=3, z=-15; \nonumber     \\
 \hspace{.5in} a=-10, b=c=-4; k=6, s=24. \nonumber
\end{eqnarray}
So far in the analysis of the possible SM embeddings, we have not
imposed the $U(1)_X$-related gauge anomaly constraints listed in
the equations of section-II.B. For any of the two embedding
solutions to give a consistent model, we need to check the
anomalies. It looks like we need a miracle to have the conditions
all just satisfied. It does {\it not} work. However, we are close.
For both solutions,  three equations, apart from the
first and last on the list, are satisfied.
This suggests a slight modification of the chiral
spectrum may get around the problem and give interesting models.
\item {\it (Model IA)}
We first point out that there is a sextet representation in $SU(4)$ that is anomaly
free. Taking the scheme $I$ solution, the simplest modification then is to
introduce a ${\bf (6,1,1,-12/5)}$. Hereafter we change the normalization
for the $U(1)_X$-charges by a factor of 5 (only for the scheme), giving all of them integral
values, for convenience ( {\it i.~e.} the sextet is then a 
${\bf (6,1,1,-12)}$, for instance). The representation  then fixes
the $[SU(4)_A]^2U(1)_X$-anomaly and  leads to three extra leptonic
singlets of  hypercharges ${\bf -3}$ and three of ${\bf -9}$ (following
the same normalization as in section-II.A). To restore
a  three family SM chiral spectrum above the EW-scale, the
vector-like partners of these singlets can be introduced as
${\bf (1,1,1,6)}$'s and ${\bf (1,1,1,18)}$'s. This happens to just
cancel the other $U(1)_X$-anomalies and gives a consistent model.
The details of the representation content of the model are shown in Table 3.
We also show all the $U(1)_X$-related anomalies as an independent
checking of the result. One interesting state here is
the pure SM-singlet ($N$)  in ${\bf (\bar{4},1,1,9)}$. If we consider
a SUSY-version of the model and promote the fermionic multiplets
to chiral supermultiplets, there is naturally a Yukawa coupling
\[      {\bf (4,3,2,5)} \hspace{.1in}{\bf (1,\bar{3},2,-14)} \hspace{.1in} {\bf (\bar{4},1,1,9)} \]
with VEV for the scalar partner of the state that gives the symmetry breaking
$SU(4)_A\otimes U(1)_X \longrightarrow SU(3)_H\otimes U(1)_Y$
as well as mass to the vector-like quark doublet($Q^{'}$). We will return to this
in the other sections below.
\item {\it (Model IIA)}
Same as in the scheme $I$ solution, we take the scheme {\it II} solution and
add a ${\bf (6,1,1,24)}$ and three ${\bf (1,1,1,12)}$'s. A
successful model is resulted. The desirable Yukawa term, analog to the one
listed above, is 
not allowed though.  Nevertheless, there is one desirable
feature for the SUSY-version which does not exist for {\it Model I}:
we have the vector-like pair of leptonic doublets identifiable with
the Higgs(ino) supermultiplets of MSSM.
\item {\it (Model IIB)}
For the scheme {\it II} solution. A $z$-value of $9$ is again  needed for
the above mentioned Yukawa term,
\[      {\bf (4,3,2,1)} \hspace{.1in}{\bf (1,\bar{3},2,-10)} \hspace{.1in} {\bf (\bar{4},1,1,9)} \]
in this case, to be admissible.
The content of singlets has then to be modified to cancel the gauge anomalies. 
The resulted model is the one we presented 
earlier\cite{728}.
\end{itemize}

We summarize the contents of the three models in Table 4, together with an extra,
minimal model, to be introduced below. Though we
noted above the possible inclusion of the symmetry breaking scalars
through supersymmetrizing, the discussion in the section  should
however be taken as addressing mainly the fermionic sector. We will
look at the possible symmetry breaking patterns and the related scalar
sectors in section-III.

\subsection{Minimal models and variations on the theme}
We can look at the model construction exercises this way: one can start
with a $SU(N)_A\otimes SU(3)_C\otimes SU(2)_L\otimes U(1)_X$ SM-like
chiral spectrum as shown in Table 1 for $N=3,4,5$ and $6$, with the
modification that two, or four, extra $\bf (1,1,2)$'s ($U(1)_X$-charge suppressed), have to be
added for the cases $N=5$ and $6$, and some singlets for all cases.
The $U(1)_X$-charges are then
determined by the possible $U(1)_Y$ hypercharges embeddings as discussed
above. Then only the $[SU(N)_A]^2U(1)_X$, the $U(1)_X$-grav.  and the $[U(1)_X]^3$
anomalies may be uncanceled. One can add in a pure, anomaly-free,
$SU(N)_A$-representation, irreducible or reducible, and adjust its
$U(1)_X$-charge(s) to cancel the $[SU(N)_A]^2U(1)_X$-anomaly. Finally, add in
singlets with the proper $U(1)_X$-charges to restore the three-family
SM chiral spectrum at the $SU(3)_C\otimes SU(2)_L\otimes U(1)_Y$ level.
All the other anomalies are canceled. This is actually a consequence of
the embedding and the fact that the SM is anomaly-free.

As an alternative \  to adding  \ extra  $SU(N)_A$-representation(s), one can also
give up the embedding of the leptonic singlets in the ${\bf (\bar{N},1,1)}$ and
simply adjust the $U(1)_X$-charge of the lattter to fix the
$[SU(N)]^2U(1)$-anomaly and proceed as above, putting in the leptonic
singlets as singlets. This actually yields models with the minimal total
number of states. The minimal model for $N=4$ scheme $I$ embedding is
listed as {\it Model Im} in Table 4, for an illustration.  
$N=5, 6$ and $3$ cases are given in
Table 5, Table 6 and Table 7 respectively. The minimal models, apart from the 
$N=3$ case, may however give SM-singlets states with unnaturally large 
hypercharges, as shown in the tables\cite{N4}.
They should be considered more as backgrounds for making modifications,
as discussed here, to obtain potentially interesting models.

Note that for each $N$, there are always two alternate hypercharge
embeddings for the quark singlets, labeled by scheme $I$ and {\it II},
as for $N=4$ shown explicitly in section-II.D. For $N=6$ (Table 6), we have
assumed that the $\bf 6$ and the $\bf \bar{6}$'s each splits into
two groups each with three  states of the same hypercharges. This
may be considered, for instance, as taking $SU(6)\longrightarrow
SU(3)\otimes SU(3)\otimes U(1)_Z$ with $U(1)_Y$ given by a linear
combination of $U(1)_Z$ and $U(1)_X$. Hence each of scheme $I$ and
{\it II} gives two possible embeddings for both the quark doublets and
leptonic  doublets, depending on which three states
are identified as the SM chiral states. The quark doublet embedding
ambiguity may be absorbed by a sign convention. Then, the two leptonic
doublet embedding identifications give two different sub-schemes for each
of scheme $I$ and {\it II}. $N=5$ has a similar situation (Table 5).
Though one group from the splitting of a ${\bf 5}$ or a ${\bf \bar{5}}$ has
two instead of three states, we allow the alternative of identifying
the two states as the SM leptonic doublets. This gives alternate models with
four extra ${\bf (1,1,2)}$, instead of two for all the other $N=5$ and $6$
schemes.

If one is willing to go into more complicated variations, one can still
temper with the $SU(2)_L$-doublet sector, by giving up the embedding
of the leptonic doublets into the ${\bf (\bar{N},1,2)}$ for instance, and get
modified models  without too much difficulty. Nevertheless, more
modifications usually introduce more states and make the pattern
deviate more from the starting theme of a SM-like chiral spectrum.

Finally, we comment on the $N=3$ case. The minimal models are shown
in Table 7. The two models,
or their modified versions, have a $SU(3)_H(\equiv SU(3)_A)$ horizontal symmetry with
the $U(1)_X$ identified directly as the hypercharge $U(1)_Y$. While
$SU(3)_H$ was among the first group  to be considered as a horizontal
symmetry for the three-family SM, the chiral fermion content,
with right-handed neutrinos, was considered to be vector-like in the
$SU(3)_H$\cite{su3}. Our models here start with a SM-like chiral
spectrum and have a very different basic structure. Whether this kind of
$SU(3)_H$ models can have a successful phenomenology we leave for
further investigation.

In the rest of the paper, we will put our
concentration  back on the $N=4$ case, which we consider most natural
in the framework and most illustrative of the general features of our approach. 
\\[.1in]

Before closing the section, we note that  a 
$SU(5)\otimes SU(3)_C\otimes SU(2)_L\otimes U(1)_Y$ model has 
recently been proposed in a different perspective\cite{FSN}. Our
approach here is partially inspired by the model. Also after completing the work
presented in this section, the author was informed about a
 $SU(2)_L\otimes SU(2)_R\otimes SU(3)_C\otimes SU(4)_G$ model\cite{PS}
which also has a partial embedding of $U(1)_Y$ into the $SU(4)_G$. The model construction
is otherwise very different from our approach. We emphasize that a
$SU(4)_A\otimes SU(3)_C\otimes SU(2)_L\otimes U(1)_X$ model from our approach
has a chiral fermion spectrum that is mostly derived from anomaly constraints 
and gives naturally three SM families as a result. This is its unique interesting feature.

\section{the quark sector and the symmetry breaking scalars }

We have presented in the previous section explicitly four 
$SU(4)_A\otimes SU(3)_C\otimes SU(2)_L\otimes U(1)_X$ models  with
SM-like chiral fermion spectra (see  also Table 4), 
from two different basic schemes of embedding the three-family SM 
(scheme $I$ and {\it II} solutions). In this section we address the important 
question of how the extra symmetries can be broken in a way that gives an experimentally
viable low-energy phenomenology. This is hooked-up with fermion mass generation
through EW-symmetry breaking. We hence discuss here also the prospect of
getting some realistic quark mass matrices. Also of interest is the possible 
phenomenology of the extra heavy quarks.

\subsection{Symmetry breaking and quark masses}
What we need, first of all, are scalars with VEVs that break
$SU(4)_A\otimes U(1)_X$ to $U(1)_Y$, with or without an intermediate 
horizontal symmetry ({\it i.\,e.} through a $G_H\otimes U(1)_Y$). Such scalars
must be  pure SM-singlet states (with  hypercharge  zero). Then we need the 
EW-Higgs doublet(s). The latter, together with the singlet
scalars, has to generate the quark masses with a nature hierarchical structure.
Particularly, the SM quark mass matrices have to be rank-one to first order,
and the extra quarks ($Q^{'}$ or $Q^{''}$) that are vector-like under SM-group
should be heavy.

So far we have not restricted our discussion to any specific model. Attentive
readers would have realized that the quark sector structure is robust within
each embedding scheme; that is still quite true even when we compare the cases
with $N=5$ and $6$ to the $N=4$ one. Also recall that the two basic embedding 
schemes differ only in the way the $\bar{u}$ and  $\bar{d}$
states are embedded.  The major difference is that the scheme $I$ models 
(e.~g.\, {\it Model IA} and {\it Im}) have an extra vector-like quark doublet 
($Q^{'}$) with electric charges $(5/3,2/3)$, while the scheme {\it II} models
(e.~g.\, {\it Model IIA} and {\it IIB}) have one ($Q^{''}$) with electric 
charges $(-1/3, -4/3)$. The former contain an extra up-type quark and the latter a
down-type one, both with the "wrong" isospin.
Then it is no surprise that the family structure of 
the scheme $I$ up-sector looks like that of the scheme {\it II} down-sector, 
and vice versa. This specific feature allows us to talk about the symmetry breaking
issue and the quark mass matrices from both schemes together. We have a 
three-quark sector and a four-quark sector; the former is the down-sector and the
latter the up-sector for scheme $I$, while their identities switch for scheme
{\it II}. The fourth quark in the four(-quark)-sector, the guy with the "wrong"
isospin,  is part of the vector-like doublet. Its  partner  is an extra quark,
of electric charge $5/3$ or $-4/3$, that does not mix with any of the others.

We again consider $SU(2)_H\otimes SU(2)_{K}\otimes U(1)_Z \subset SU(4)_A$
and use the quantum numbers in the subgroup embedding as a convenient
label for the states. For instance, the ${\bf (\bar{4},1,1,9)}$ contains
the four states denoted by ${\bf (2_{\pm},1)_{1,9}}$ and  
${\bf (1,2_{\pm})_{-1,9}}$ where the  ${\bf (1,2_{+})_{-1,9}}$ state has 
$T_{3(K)}=+1/2$ and $U(1)_Z=-1$. ( $U(1)_X=9$; here in the notation  we suppress the
$SU(3)_C$ and $SU(2)_L$ quantum numbers.) From the results of section-II.D, 
the hypercharges are then given by 
\begin{equation}
U(1)_Y = \pm \left ( \frac{1}{2} U(1)_X + \frac{3}{2} \left [ U(1)_Z - 4T_{3(K)} \right ] \right )
\end{equation}
with the positive and negative signs corresponding to the    
scheme $I$ (with above modified $U(1)_X$ normalization) and {\it II} solutions  
respectively. Our example state  ${\bf (1,2_{+})_{-1,9}}$ is then a state of
hypercharge zero. Also, the Higgs doublets that are capable of giving masses
to the SM-quarks satisfy the equation
\begin{equation}
\frac{1}{2} U(1)_X + \frac{3}{2} \left [ U(1)_Z - 4T_{3(K)} \right ] = \pm 3
\end{equation}
with the positive and negative signs correspond to the ones coupling to the  four-
and three(-quark)-sectors respectively, independent of which schemes or models
we are talking about. We can now look at the particular Higgs-VEVs needed to 
generate mass for each entries of the quark mass matrices of both sectors.
They are given by
\begin{equation}
$$M^{(3)} \sim {\bf \left(\begin{array}{ccc} 
{\bf \left \langle (3_0,1)_{0,-6}\right \rangle } &  &  \\
{\it or} & {\bf \left \langle (3_+,1)_{0,-6}\right \rangle } 
& {\bf \left \langle (2_+,2_-)_{2,-6}\right \rangle } \\
{\bf \left \langle (1,1)_{0,-6}\right \rangle }   &  &  \\
 & & \\
  & {\bf \left \langle (3_0,1)_{0,-6}\right \rangle } &  \\ 
{\bf \left \langle (3_-,1)_{0,-6}\right \rangle } &  {\it or} 
&  {\bf \left \langle (2_-,2_-)_{2,-6}\right \rangle } \\
  & {\bf \left \langle (1,1)_{0,-6}\right \rangle } &  \\
 & & \\
  &  &   {\bf \left \langle (1,3_0)_{0,-6}\right \rangle }  \\
{\bf \left \langle (2_-,2_+)_{-2,-6}\right \rangle } 
& {\bf \left \langle (2_+,2_+)_{-2,-6}\right \rangle } &  {\it or} \\
  &  &   {\bf \left \langle (1,1)_{0,-6}\right \rangle }
\end{array}\right) }$$ 
\end{equation}
and 
\begin{equation}
$$M^{(4)} \sim {\bf \left(\begin{array}{cccc}
{\bf \left \langle (2_-,1)_{1,3}\right \rangle } & {\bf \left \langle (2_+,1)_{1,3}\right \rangle } 
& {\bf \left \langle (1,2_-)_{-1,3}\right \rangle } & {\bf \left \langle (1,2_+)_{-1,3}\right \rangle } \\
 & &  & \\
{\bf \left \langle (2_-,1)_{1,3}\right \rangle } & {\bf \left \langle (2_+,1)_{1,3}\right \rangle }
& {\bf \left \langle (1,2_-)_{-1,3}\right \rangle } & {\bf \left \langle (1,2_+)_{-1,3}\right \rangle } \\ 
 & &  & \\
 &  &  & {\bf \left \langle (1,3_0)_{0,-6}\right \rangle }   \\ 
 {\bf \left \langle (2_-,2_-)_{2,-6}\right \rangle }  &  {\bf \left \langle (2_+,2_-)_{2,-6}\right \rangle } 
 & {\bf \left \langle (1,3_-)_{0,-6}\right \rangle } &   {\it or} \\
 &  & & {\bf \left \langle (1,1)_{0,-6}\right \rangle }   \\
 & &  & \\
0 &  0 & 0 & {\bf \left \langle (1,2_+)_{-1,9}\right \rangle } 
\end{array}\right) }$$ ,
\end{equation}\\[.1in]
where we noted that the ${\bf 3_+, 3_0}$ and ${\bf 3_-}$ correspond to states in
a $SU(2)$ triplet with $T_3 = +1, 0$ and $-1$ respectively. As remarked above, the VEV 
${\bf \left \langle (1,2_+)_{-1,9}\right \rangle }$ is in a pure SM-singlet state
and, if admitted, results in the symmetry breaking 
$SU(4)_A \longrightarrow SU(3)_H\otimes U(1)_{Z^{'}}$ and give masses to both
quarks in the vector-like doublet ($Q^{'}$ or $Q^{''}$). The mass term  corresponds to
the $44-$entry of the $M^{(4)}$-matrix. With only this VEV,  
$SU(3)_H$ then serves as an unbroken horizontal
symmetry. The VEV is hence a desirable one, though not necessarily the  $SU(3)_H$.
All the other VEVs shown in the mass matrices are in the zero electric charge
components of doublets(EW). The zero matrix-entries correspond to states of zero
electric charges that would only be available from $SU(2)_L$ triplets 
which we assume nonexisting. Note that the entries of the first two rows of $M^{(4)}$
are identical, giving one zero-eigenvalue. One naturally very small quark mass
eigenstate is therefore to be expected.

Assuming an un-broken $U(1)_Y$, the VEV in ${\bf (1,2_+)_{-1,9}}$ is the only one
admissible in $\phi_0 = {\bf (\bar{4},1,1,9)}$. Actually, we can switch the argument
the other way round and use the natural VEV
\begin{equation}
$$\left \langle \phi_0 \right \rangle =\left(\begin{array}{cccc}
0 & 0 & 0 & v
\end{array}\right) $$ 
\end{equation}
to define the remnant $SU(3)_H$ and $U(1)_Y$ symmetry.
Other representations may then have more zero hypercharge states. 
The interesting thing is: all the doublet VEVs in $M^{(3)}$ and $M^{(4)}$  listed above
can come from just two representations, a ${\bf (15,1,2,-6)}$ and a
${\bf (\bar{4},1,2,3)}$. One easy way to obtain rank-one EW-quark
mass matrices is to  take just the scalar $\Phi = {\bf (15,1,2,-6)}$ and  give VEVs 
only to the  ${\bf (1,3_0)_{0,-6}}$ and ${\bf (1,3_-)_{0,-6}}$ states. The VEVs actually
perserves the $SU(2)_H$ as a horizontal symmetry which keeps the lighter
two families massless. The abundance of neutral scalars coupling to
the quarks, here from a single Yukawa term involving  $\Phi$, leads to worries about
FCNC constraints\cite{fcsc}, which basically requires most if not all the other
scalar in the  $\Phi$ without a VEV to be heavy ($\geq 200 TeV$).  But there is a 
way that this can be done.  A ${\bf (\bar{4},1,1,-3)}$ gives three zero
hypercharge states, in an anti-triplet of the $SU(3)_H$. If we take two such scalar
multiplets, denoted by $\phi_a$ ($a=1$ or $2$), with natural VEVs
\begin{equation}
$$\left \langle \phi_1 \right \rangle =\left(\begin{array}{cccc}
v_1 & 0 & 0 & 0
\end{array}\right) $$ ,
\hspace{.5in}
$$\left \langle \phi_2 \right \rangle =\left(\begin{array}{cccc}
v_1^{'} & v_2 & 0 & 0
\end{array}\right) $$ ,
\end{equation}
a mass term for   $\Phi$   of the form
\[
C_{ab} \phi_{ai} \phi_b^{\dag  j} \Phi^k_j \Phi^{\dag i}_k
\]
gives masses to all components of the scalar bearing non-trivial $SU(2)_H$
quantum numbers. This is basically the mass term used previously in a $SU(3)_H$ 
model\cite {SH}. It can be interpreted as enforcing the coupling of 
$\phi_{a} \phi^{\dag}_b$, with  their nonzero VEVs only in the $SU(2)_H$
nontrivial directions, to the $\Phi \Phi^{\dag}$ only through the part
that transforms as a ${\bf (3,1)_0}$. Recall the splitting
\[ 
{\bf 15 \longrightarrow (1,1)_0 + (1,3)_0 + (3,1)_0 + (2,2)_{2} + (2,2)_{-2}}. \]
It can then be seen easily that the term splits the multiplet and leaves only
the ${\bf (1,3)_0}$ states massless, with EW-scale masses then assumed to be
generated by other mechanisms, for instance radiatively.
The sub-multiplet, apart from the $\Phi _{+} ={\bf (1,3_+)_{0,-6}}$ doublet 
which contains no neutral states, gives the EW-breaking doublets we want.  The 
$\Phi _{+}$ doublet has scalar states with electric charges $(\mp 1,\mp 2)$\cite{sign}
giving Yukawa couplings $\bar{d} \Phi_+ Q^{'}$ or $\bar{u} \Phi_+ Q^{''}$. The 
doubly-charged scalar state in particular can be considered a 
novel part of the prediction from our  models.

It is interesting to see that one can go on to construct almost the full quark mass
matrices even without introducing further extra scalars. For instance, combining 
$\Phi$ and $\phi_0$, through a dimension-5 term can give $1 / M_0 $ suppressed 
effective mass term of the form
\[
 \left \langle \Phi  \right \rangle v_0 / M_0  \]
to four more entries in $M^{(4)}$ as shown in Table 8a, leading to a second 
non-zero SM quark mass
eigenvalue. Here $M_0 $ would have to be some mass scale higher than $v_0$.
To name two possibilities: it can be $M_{Pl}$ with the effective Yukawa coupling 
being gravitationally generated, if $v_0$ is at a very high scale\cite{gs}; or it can 
be the mass scale of some other vector-like fermions with the effective coupling 
being generated {\it a la} Froggatt-Nielsen\cite{fn} (see Fig. 1 for an illustration).
In a non-SUSY-compatible scenario, further combining with
 $\phi_a (a=1,2)$ can give at least a second mass eigenvalue for  $M^{(3)}$ (see 
Table 8a). For the scheme $I$ models, this corresponds to a natural mass hierarchy
\[  m_t  ,   m_b  >  m_c  >   m_s >  m_d , m_u \]
leaving only the $m_t/m_b$ ratio to be fixed by their VEVs and the very small 
$m_d$ and $m_u$ to be generated by radiative mechanism. There is also an 
alternative but similar possibility of starting with   $\Phi = {\bf (\bar{4},1,2,3)}$
and the same set of singlet scalar as shown in Table 8b.  
For the scheme $I$ models, this corresponds to a natural mass hierarchy
\[  m_t  >   m_b  ,  m_c  ,   m_s >  m_d , m_u \ \ . \]
 
We consider the above analysis a partial success in generating the quark mass
hierarchy and an illustration of the promising potentials of our approach. To complete
the quark mass matrices construction, radiative mass generations have to be analyzed
and explicit mass scales have to be fixed.  Other scalar VEVs 
may also be introduced. This should be done within the context of a specific model.

\subsection{The heavy quarks}
Recall that our approach gives a characteristic extra quark doublet
 ($Q^{'}$ or $Q^{''}$) which is vector-like
under the SM group. In the discussions above, we adopt a  symmetry-breaking 
scalar $\phi_0 = {\bf (\bar{4},1,1,9)}$ which gives masses  $M_Q$ to the quark doublet 
through a Yukawa term. If the mass is at a high scale, as is to be expected for the
symmetry breaking ({\it i.\,e.\,} $v_0\geq 200TeV$), there would hardly be any
interesting  accessible phenomenology. However, if the mass for the quarks were
substantiallly below symmetry breaking scale, perhaps due to a small (effective)
Yukawa coupling as in the cases of the SM quarks apart from the top, it could have
a much more interesting experimental implication. In this subsection, we discuss
some of the interesting possibilities under the assumption.

The two constituents of  the doublet behave differently: one is the fourth up- or
down-type quark ($q_4$), which most propably mixes with the other light quarks in
$M^{(4)}$; the other is an exotic guy, an "above-the-top" or "below-the-bottom"
quark ($q_*$), with electric charge $5/3$ or $-4/3$ respectively. The two scenarios
correspond to the two schemes of embedding; and the identity of this extra quark
doublet is what distinguishes models of the two schemes. 

The heavier of the two quarks, $q_4$ or $q_*$, can decay into the lighter and
a $W$-boson, if their mass splitting is large enough. However, large mass splitting
is unfavorable, from both theoretical and experimental perspectives. From the
group structure of the fermion spectrum, we expect a mass degeneracy between
the two quark states, only to be lifted by the small mixing between  $q_4$ and
the light quarks. The mass splitting is also constrained by the experimental limit
on its contribution to the $\rho$-parameter. In a degenerate scenario, the major 
decay mode for $q_*$ is likely to involve Yukawa vertices, as the quark is 
assumed to be much lighter than the extra gauge bosons. For instance, if 
 $\Phi = {\bf (15,1,2,-6)}$ is taken to provide the EW-Higgses, a light doublet
of charged Higgses is predicted and both $q_*$ and  $q_4$ can decay into both an
up- or down-type quark with one of the Higgses.  Under the situation of small
Yukawa couplings or nonexistence of such scalars,  $q_*$ is likely to be relatively
stable, while  $q_4$ can still decay through Yukawa couplings responsible for 
its mass mixing.

If the quarks were not too fast-decaying, they may, upon QCD-confinement, 
form new mesonic and baryonic states, among themselves or with the light
quarks. All these QCD-singlet states have integral charges. The states involving
$q_*$ would be more interesting; for example, there could be doubly-charged
mesons.

Phenomenology of SM vector-like quarks has been studied\cite{4q} particularly
extensively from the perspective of spontaneous CP violation\cite{4qcp}, and
recently in the context of fixing the $R_b-R_c$ anomaly\cite{rbc}. The type of
analysis is particularly relevant for the quark $q_4$.  CP violation has also been
ascribed to horizontal  interactions by other authors\cite{cpv}. The extra gauge
symmetries in our approach has some similarities to horizontal symmetries, while 
our models also have the vector-like quarks. CP nonconservation properties of
models from our approach would be very interesting. We will, however, leave the
issue to further investigation. We discuss here FCNC constraints and possible
impact on  the $R_b-R_c$ anomaly. Most of the result can be easily adopted  from the
cited references.

The kind of Higgs configurations for quark mass generation we discussed above
has basically the Natural Flavor Conservation feature in it at the EW-scale. 
However, the mass mixing involving   $q_4$ in $M^{(4)}$ does generate FCNC's
at the $Z$-vertex, essentially because  of the "wrong" isospin of  $q_4$. Taking
a biunitary rotation to diagonalize the first $3\times 3$ block, we can write
\begin{eqnarray}
\tilde{M}^{(4)} 
=   \left( \begin{array}{cc}
U_R^{\dag}  &  \\ 
        &     1 
\end{array}  \right)  M^{(4)}  \left( \begin{array}{cc}
U_L  &  \\ 
        &     1 
\end{array} \right)   \nonumber  \\
\sim   \left( \begin{array}{cccc}
m_1 &  &  & x_1 \\
  & m_2 & & x_2 \\
 &  &  m_3 & x_3 \\
0 & 0 & 0 & M_Q
\end{array} \right) \ .
\end{eqnarray}
In terms of the mass eigenstates $q^{'i}$
\begin{equation}
L_Z^{FCNC} = \beta_{ij} \bar{q}_R^{'i} \gamma_{\mu} q_R^{'j} Z^{\mu}
\end{equation}
for $i \neq j$, where
\begin{equation}
 \beta_{ij} = \left (\mp \frac{1}{2} \right ) \frac{g_2}{cos\theta_W} (K_R)^*_{4i}  (K_R)_{4j} 
\end{equation}
with $K_R^{\dag} \tilde{M}^{(4)} K_L = {\it diag}\{ M^{(4)} \}$ and the signs
correspond to the two schemes\cite{sign}.
Here the more interesting situation comes from scheme {\it II} embedding
where the four-sector is the down-sector. There, a stronger constraint coming
from Kaon decay $K_L\longrightarrow \mu^+ \mu^-$ search requires
$ \beta_{12} < 10^{-6}$\cite{rK}. Now 
\begin{equation} 
(K_R)^*_{4i} = x_i/M_Q = (U_R^{\dag})_{ij} M^{(4)}_{j4} / M_Q ;
\end{equation}
and from the group symmetry structure of $M^{(4)}$ we expect then
\begin{equation}
x_1 \sim x_2 < 10^{-3} M_Q .
\end{equation} 
Putting $x_2 = m_s$ implies only $M_Q > 150GeV$. 
Corresponding FCNC contributions from the left-handed component is much
further suppressed as 
\begin{equation} 
(K_L)^*_{4i} = \frac{x_i}{M_Q} \frac{m_i}{M_Q} .
\end{equation}
$x_3$, however, is expected to be larger. Actually, the group symmetry structure
gives
\begin{equation}
x_3 = - M^{(3)}_{33}
\end{equation}
which, in the case of scheme {\it II} models, gives
\begin{equation}
x_3 \sim m_t .
\end{equation}
The mixing serve in the right direction to fix the $R_b$-anomaly.  Quantitatively,
we need
\begin{equation}
\left ( \frac{x_3}{M_Q} \right )^2 = 0.059 \pm 0.016 ,
\end{equation}
giving a value of $M_Q$, or rather $m_{q_4}$ in particular, to be
\begin{equation}
m_{q_4} \sim 635-840 GeV .
\end{equation}

In the scheme $I$ models, mixing with $q_4$ goes to the up-sector. FCNC constraints
are much less severe. However, $x_2$ can be used to  reduce $R_c$ only in the context
of large mixing, and $x_3$ can be used to give large mixing with the top which
could then reduce $R_b$ through changing the top-loop effect\cite{BBCLN}. The
scenario flavors a much lighter $q_4$, perhaps below $m_t$ pulling the ratio 
between $M_Q$ and the symmetry breaking scale much smaller. That makes 
it very unlikely to fit in with other phenomenological aspect for a complete model
from our approach and  therefore less attractive.

\section{The gauge sector, and more}

\subsection{The gauge sector}
The gauge bosons of $SU(4)_A$ are contained in an adjoint ${\bf 15}$, nine among them 
are neutral. One among the nine, the one that transforms as ${\bf (1,1)_0}$ in
the case of a simple 
 $SU(4)_A \longrightarrow SU(2)_H\otimes SU(2)_K\otimes U(1)_Z$ symmetry
breaking, mixes with the $U(1)_X$-boson to give
an heavy state and a massless one. The latter is to be identified as the $U(1)_Y$-boson
when the symmetry is broken to that of the SM. The heavy state together
with the other eight, which also develop heavy masses,  behave very much like 
gauge bosons of horizontal intereactions and have to satisfy 
similar FCNC constraints\cite{gs1,gs2}. The other six states of the adjoint ${\bf 15}$
are states of non-zero $T_{3(K)}$ and correspond to charged gauge-bosons:
 ${\bf (1,3_-)_0}$ and  ${\bf (2_{\pm},2_-)_2}$ of charge $\pm 1$, and their
conjugates
 ${\bf (1,3_+)_0}$ and  ${\bf (2_{\pm},2_+)_-2}$ of charge $\mp 1$\cite{sign}.
Such gauge bosons are {\it  not} to be expected in a horizontal symmetry framework. The 
charged gauge bosons  can contribute to FCNC only through loop-diagrams.
As they are expected to have masses at the same scale with the neutral ones, such
contribution would play a secondary role.

The strongest FCNC constraint relevant is given by the process 
$K_L\longrightarrow \mu e$ which gives lower bound on the neutral gauge boson
mass\cite{rK} as
\begin{equation}
M_{\mathcal{X}} \sim 220 TeV
\left [ \frac{10^{-12}}{B(K_L \rightarrow \mu e)}\right ]^{1/4} ,
\end{equation}
comparable to that on the heavy neutral scalars discussed in the  section-III.A
above. A potentially stronger bound coming from $K_L - K_S$ mass difference
is liable to a reduction factor of $\Delta_{M_{\mathcal{X}}}/ M_{\mathcal{X}}$
where the numerator denote the lack of degeneracy among the gauge boson
masses\cite{gs1}.

The extra charged gauge bosons give  extra contributions to charge 
current processes with special characteristic.  For instance, for the SM quarks,
the couplings involve only a $\bar{u}$ and a $\bar{d}$  in the same 
${\bf (\bar{4},\bar{3},1,x)}$ multiplet (see Fig.2); with  the type of Higgs structure 
discussed above, at least one of them would be predominately a third family
guy. The processes are of course suppressed by the same gauge boson
mass scale $M_{\mathcal{X}}$.

Another interesting aspect concerning the gauge sector is the RG-runnings
of the gauge couplings. While there is no obvious gauge group unification,
string-type gauge coupling unification may not be ruled out. The RG-runnings
consideration is likely to  impose limits on the acceptable mass scales involved.
One particularly interesting aspect under the assumption of gauge coupling
unification is that  if the $SU(4)_A$ is asymptotically free, to get the correct 
count of the number of families,  we would have to 
make sure that it breaks before it confines.  With a strong asymptotic freedom, 
this may even set a pretty high lower limit on the symmetry breaking
scale, making the extra vector-like fermions such as the quark doublet 
($Q^{'}$ or $Q^{''}$) less likely to be accessible to low-energy phenomenology.
The RG-runnings are of course dependent on the specific details of the models 
including the full content of the scalar multiplets, which may include extra 
ones needed to give masses to some of the leptonic sector states. We are hence
going to just sketch briefly some possible scenarios.

As an example, taking a supersymmetrized version of the spectrum of {\it Model IIB},
without extra scalar, the coefficients for the first order $\beta$-functions 
are given by
\begin{equation}
 (b_4,b_3,b_2,b_1)=\frac{1}{16\pi^2}(-5,-1,4,233/24)
\end{equation}
where we have normalized the $U(1)_X$-charge by 1/24. We are actually 
interested mainly in the  coefficients  $b_3$ and   $b_4$.  
Firstly, we can see that the model  maintains 
the $SU(3)_C$ asymptotic freedom. The coefficient $b_3$ is actually very robust.
It is universal for all the models and would not change up any modification of
the models as suggested in section-II.E. Without SUSY, and without colored scalars,
 $b_3$ will only be more negative the asymptotic freedom stronger.
$SU(4)_A$ asymptotic freedom  looks uncomfortably strong.
However, for any of the models to be realistic, we very likely  have to put in 
extra scalar multiplets to take care of the various symmetry breakings. This almost
necessarily changes all the coefficients except  $b_3$. Following our discussion
on quark mass generation in section-III.A, we consider taking  
$\Phi = {\bf (15,1,2,-6)}$ with  the singlet scalars $\phi_1$ and $\phi_2$ as  extra
supermultiplets. This exactly kills the asymptotic freedom and gives  $b_4=0$.
But extra supermultiplets are then needed to cancel anomaly contributions from
their fermionic partners. Hence we can see that without SUSY, $SU(4)_A$
is likely to have strong asymptotic freedom; while with SUSY and  
$\Phi = {\bf (15,1,2,-6)}$, it is likely to  lose it totally. No conclusive statement
could be made either way without a specific detailed model. 
Scalars like $\Phi = {\bf (15,1,2,-6)}$ also have a large effect on the $b_2$ coefficient
and hence the scale where the  $SU(3)_C$ and  $SU(2)_L$ couplings meet. The
possible perturbative limit on the $SU(4)_A$ coupling is hence very model dependent.

\subsection{The leptonic sector}
Our approach in general yields a leptonic sector richer in content than the quark
sector. Unlike the latter, the former  is where the flexibilities in fixing
the detailed fermion spectrum lie.
Scheme $I$ embedding gives basically an extra vector-like leptonic doublet
($L^{'}$) of hypercharge ${\bf -9}$, or constituent states of electric charges
one and two. Again, for scheme {\it II} embedding, the extra doublet is  just a vector-like
version of the SM ones. The  list of vector-like leptonic singlets
is very model dependent. They come in charge zero, one or two for three of
our explicit models listed in Table 4, while the only exception is {\it Model IA}
which has charge $1/2$ and $3/2$ singlets\cite{N4}.  A complete formulation of the
symmetry breaking has to generate a realistic mass spectrum for the chiral and
vector-like leptons too. This however would not be possible without fixing the
contents of the sector.

We want to note about two interesting features. The first one is that extra
vector-like states with the same quantum numbers as the SM leptonic singlets 
($E \equiv e^+$) is very common. They are in three of the explicit models, except 
{\it Model IA}. If there is no direct mass term for some or all of the SM 
charged leptons and they get mass only through mixing with the heavy vector-like
states, the smallness of the charged lepton masses could have a natural 
explanation through a seesaw type mechanism.

The second feature of interest is the existence of right-handed neutrino states($N$)
in the our SM-like chiral spectra. For example, there is one such state in both 
{\it Model IA} and {\it Model IIB}, and three of them in {\it Model IIA}.  
Right-handed neutrinos can of course have Majorana masses invariant under the
SM symmetry. Their existence can lead to desirable small neutrino masses
through the seesaw mechanism. In an early work on horizontal symmetry\cite{su3n},
it has been suggested that the right-handed neutrino mass scale is to be
identified as the  breaking scale of the extra (horizontal) symmetry. This 
suggestion could work more naturally in the models from our approach as
we have fermion spectra that are full chiral (SM-like) yet able to yield states
that are to be identified with right-handed neutrinos at the SM symmetry level.
Moreover, these states come out in our models as a result, without their
existence being assumed beforehand. Actually, they are bound to exist in any
SUSY version of models from our approach, for their superpartners are the
zero hypercharge scalars whose VEVs are  needed for the symmetry 
breaking. And from the discussion in section-III.A, we can see that at least a
few of them are likely to be needed for a realistic chiral fermion mass generation.
The scale of right-handed neutrino Majorana masses is expected to be around
$10^{13}GeV$, in a simple see-saw picture. Whether the scale is compatible with the constraints from
the issues related to the RG-running of the gauge coupling is questionable. And 
as remarked above, such a high symmetry breaking scale almost neccessarily
imply that the extra vector-like states would not give much interesting and
accessible phenomenology.

 We also note that effective Majorana mass for
(left-handed) neutrinos from higher dimensional terms have been studied
more recently from the horizontal symmetry framework\cite{BCS}. The type
of mechanism as an alternative for neutrino mass generation may also be
relevant for models from our approach.

The right-handed neutrino states from our approach have a very interesting
characteristic feature: they can decay into a charged lepton and a meson,
for example $N\longrightarrow e^+K^-$, through one of the extra charged 
gauge bosons as shown in Fig.3. As mentioned earlier in relation to the gauge
bosons involved, at least one of the quarks in the charged meson resulted would
be predominately a third family guy (see Fig.2). But any particular mass 
eigenstate could involved through the mass mixing. In most of the cases, 
the charged lepton is identifiable with the chiral leptons also only through
mass mixing. Other similar decay modes might be possible through Yukawa
vertices instead of gauge ones.

\subsection{Incoporating SUSY?}
Supersymmetrizing the SM is a popular way to stablize the EW-scale. Our approach
to the family structure, with the extra symmetries broken through Higgs
mechanism and elementary EW-Higgs doublets, is in need of SUSY, or an
alternative mechanism, to stablize the gauge hierarchy. Though the one-family
SM has the very elegant chiral fermion spectrum, the Higgses needed for the
symmetry breaking have to be additionally postulated. In the supersymmetrized
theory, however, scalars and fermions are in general on pretty equal footing; they are
partners in chiral supermultiplets. The fermionic partner of each Higgs, for
example, contributes  to the gauge anomalies. Our SM-like 
chiral fermion spectra are "derivable" basically from gauge anomaly constraints
in a way similar to the one-family SM. A supersymmetrized version of any of
such models would be really self-contained if the needed symmetry-breaking
or mass-generating scalars (Higgses) are available in the chiral spectrum of
then supermultiplets. A realistic theory of the type has to account for all
the scalar masses, as well as fermion masses, may be with soft SUSY-breaking
terms included. This is a very ambitious goal that we are far from achieving here;
nor are we sure that it can be done  within our approach.
We just want to highlight some of the potentials of our models in the perspective.

In section-II.D, we have already remarked that the  ${\bf (\bar{4},1,1,9)}$
in {\it Model IA} and  {\it Model IIB}, promoted to a supermultiplet, naturally
incorporates the scalar ($\phi_0$ in section-III.A) that "extracts" $U(1)_Y$ out 
of the extra symmetries and gives masses to the extra quarks ($Q^{'}$ or
$Q^{''}$). This part falls in line with the goal easily. Then in   {\it Model IIA} and 
{\it Model IIB}, the  ${\bf (\bar{4},1,2,3)}$,
together with the ${\bf (1,1,2,6)}$, could be identified as containing both the
SM leptonic doublets and the Higgs/Higgsino doublets. If that works,, {\it Model IIB} could
be really self-contained, as it  is actually
constructed with that as the motivation. Now, how realistic could this be?

First thing we have to note is that when both the SM lepton/slepton doublets and the
Higgs/Higgsino doublets are identified from the  ${\bf (\bar{4},1,2,3)}$
and the  ${\bf (1,1,2,6)}$, we do not
need to have the big split in mass among the ${\bf \bar{4}}$ components as discussed
in section-III.A. In fact, all parts of the  multiplet would have to have masses at  
EW-scale or lower. The scalar masses can be obtained from the soft SUSY-breaking
terms. But then there are also the $\mu$-terms($\mu_i L_i\bar{L}$) which
lead to mixing among the states, and the fermions among the $L_i$'s have to be 
identified as leptons and Higgsinos afer a EW-scale matter-Higgs rotation\cite{mH}.
The phenomenological implications of the type of mixing have recent analyzed by 
various authors recently\cite{mH,mH2}, mainly from the perspective of MSSM. 
They are: violation of lepton-number in the related quark Yukawa couplings; generation
of sneutrino VEV(s); mass generation for neutrino(s) through mixing with a gaugino or
a Higgsino. The first one can lead to FCNC's that are potentially dangerous. However,
lepton-number conserving and violating Yukawa couplings are shown to be diagonalized
simultaneously, suppressing the problem. And with a alignment between the $\mu_i$
and the sneutrino VEVs, the resultant neutrino mass(es) are shown to be acceptable. 
To use the above mentioned result in our model and piece together a consistent 
picture, the details of the quark and lepton mass generations, and the scalar masses including the soft SUSY-breaking parts have to be analyzed within the framework of
our extended symmetry. The scenario  worths pursuing for its great 
esthetic appeal and is under investigation.

Given the flexibility in modifying the exact spectrum discussed in section-II.E,
an alternative way to  incorporate SUSY  is to find a anomaly-free
chiral supermutiplet spectrum that incorporates also the needed scalars, 
for example the ${\bf (15,1,2,-6)}$ and others.  This is
likely to deviate from our starting theme of a SM-like chiral spectrum but might
yield interesting models. 

The last resort is to put in the needed scalars in vector-like pairs of the full 
symmetry. Or, giving up SUSY, one will have to find other workable mechanism
of dynamical mass generation that is compatible with the structure of our
approach and with all the experimental constraints.

\section{concluding remarks}
We have elaborated on our  new approach to the family structure of the SM 
proposed earlier\cite{728}. The unique feature of our approach is to start
with a SM-like chiral fermion spectrum largely "derivable" from gauge 
anomaly constraints within an extended symmetry of 
$SU(N)\otimes SU(3)\otimes SU(2)\otimes U(1)$.
The most suggestive $N=4$ case also naturally yields SM-embeddings{\em  with
the number of families  being three as a result}. Model construction with 
$N=3, 5$ and $6$ have also been discussed. The approach has
some flexibility in fixing the exact details of the spectrum, mainly in the leptonic
sector. We have presented four explicit $N=4$ models from two schemes
of embedding the SM and used them to address
some of the most interesting phenomenological features, concentrating on
those that are more general to the approach than the more model specific ones. 

The quark-sector is quite model independent, though differs between the
two embedding schemes. With the use of a relative simple
set of scalar multiplets, we have illustrated a  possible symmetry breaking
pattern that can evade the stringent FCNC constraints and naturally gives
SM quark mass matrices with a hierarchical structure. The analysis flavors
the scheme $I$ models a bit more. Other interesting 
predictions from the scenario include: a doubly-charged Higgs at EW-scale; 
a new vector-like quark doublet of an extra up-type and a above-the-top quark,
or an extra down-type and a below-the-bottom quark, depending on the embedding
scheme, possibly with relatively accessible masses; in scheme {\it II} models,
assuming the extra down-type quark mix with the bottom to just fix the observed
$R_b$-anomaly gives the mass of the extra quarks to be around $715GeV$;
existence of six extra charged gauge bosons; more charged leptonic singlets
depending on model specifics;  and likely right-handed neutrino states with 
interesting characteristic decay modes through the charged gauge boson channels.
The RG-running of the $SU(4)_A$ coupling can go either way depending on the
details of the model but asymptotic freedom for $SU(3)_C$ is always maintained.

A SUSY version of {\it Model IIB} could be a really self-contained model, if the 
Higgses are taken to be in the   ${\bf (\bar{4},1,2,3)}$ supermultiplet together with
the lepton doublets. More work is needed to see if the model could then be made
consistent and realistic. 

Finally, we note that we have discussed in the paper a few interesting 
phenomenological possibilities that are not necessarily compatible with one 
another. Further investigation, particularly in the context of a specific model,
is needed to see which ones among them can be fitted together into a single
consistent model extending the SM and solving the family problem, at least
theoretically.

\acknowledgements
The author wants to thank P.H.Frampton for encouragement and support,
and for valuable discussions.

This work was supported in part by the U.S. Department of
Energy under Grant DE-FG05-85ER-40219, Task B.\\

\clearpage

{\bf Table Caption.}\\

Table 1: Suggestive representation structures from
the standard model to
$SU(N)_A\otimes SU(3)_C\otimes SU(2)_L\otimes U(1)_X$
with three quark families. Note that for $N=3$,
we have a natural trivial embedding, $U(1)_X\equiv U(1)_Y$.

\bigskip

Table 2: Embedding the three family SM.

\bigskip

Table 3:  {\it Model IA} -- representation structures, anomaly cancellations
and the hypercharge embedding.

\bigskip

Table 4: Explicit contents of four $SU(4)_A\otimes SU(3)_C\otimes SU(2)_L\otimes U(1)_X$ models.

\bigskip

Table 5: Minimal  $SU(5)_A\otimes SU(3)_C\otimes SU(2)_L\otimes U(1)_X$
models.

\bigskip

Table 6: Minimal  $SU(6)_A\otimes SU(3)_C\otimes SU(2)_L\otimes U(1)_X$
models.

\bigskip

Table 7: Minimal  $SU(3)_H\otimes SU(3)_C\otimes SU(2)_L\otimes U(1)_Y$
models.

\bigskip

Table 8a: A possible pattern of mass generation, starting with  $\Phi = {\bf (15,1,2,-6)}$.

\smallskip

Table 8b:  An alternative possible pattern of mass generation,  starting with  $\Phi = {\bf (\bar{4},1,2,3)}$.

\bigskip
\bigskip

{\bf Figure Caption.}\\

Figure 1: Illustrations of Froggatt-Nielsen mechanism for quark mass generation.
Scheme $I$ quarks are used. Only $SU(4)_A$ and $U(1)_X$ quantum numbers
shown. Two possible tree graphs that can generate the $M^{(4)}_{13},
M^{(4)}_{23}, M^{(4)}_{14}$ and $M^{(4)}_{24}$ at the second level (see 
Table 8a) illustrated.

\bigskip

Figure 2: Charged gauge boson vertices with quark singlets. Note that the quark
line switch identity from a $\bar{u}_3$ to a one of the three $\bar{d}$ for scheme
$I$ models; while for scheme  {\it II} models from  $\bar{d}_3$ to a  $\bar{u}$.
Similar vertices involving the quark doublets always involves the $Q^{'}$ or
$Q^{''}$.

\bigskip

Figure 3: Characteristic decay mode of a right-handed neutrino. Note that we
have $i=3$ or $j=3$ for scheme $I$ or {\it II} models respectively,
assuming  $\Phi = {\bf (15,1,2,-6)}$.

\clearpage
\footnotesize

Table 1. Suggestive representation structures from
the standard model to
$SU(N)_A\otimes SU(3)_C\otimes SU(2)_L\otimes U(1)_X$
with three quark families. 

\vspace{.1in}

\noindent
\begin{tabular}{|c||c|c|c|c|c|}
\hline\hline
 SM                             & $N=3$                         & $N={\bf 4}$                           & $N=5$                         & $N=6$                 & $N=7$         \\
\hline\hline
				& ${\bf (3,3,2)}$               & ${\bf (4,3,2)}$               & ${\bf (5,3,2)}$               & ${\bf (6,3,2)}$               & ${\bf (7,3,2)}$               \\
				& ${\bf (\bar{3},\bar{3},1)}$   & ${\bf (\bar{4},\bar{3},1)}$   & ${\bf (\bar{5},\bar{3},1)}$   & ${\bf (\bar{6},\bar{3},1)}$   & ${\bf (\bar{7},\bar{3},1)}$   \\
				& ${\bf (\bar{3},1,2)}$         & ${\bf (\bar{4},1,2)}$         & ${\bf (\bar{5},1,2)}$         & ${\bf (\bar{6},1,2)}$         & ${\bf (\bar{7},1,2)}$         \\
				& ${\bf (\bar{3},1,1)}$         & ${\bf (\bar{4},1,1)}$         & ${\bf (\bar{5},1,1)}$         & ${\bf (\bar{6},1,1)}$         & ${\bf (\bar{7},1,1)}$         \\
\hline
${\bf (3,2)}$                   &                               & ${\bf (1,\bar{3},2)}$         & ${\bf (1,\bar{3},2)}$         & ${\bf (1,\bar{3},2)}$         & ${\bf (1,\bar{3},2)}$         \\
				&                               &                               & ${\bf (1,\bar{3},2)}$         & ${\bf (1,\bar{3},2)}$         & ${\bf (1,\bar{3},2)}$         \\
				&                               &                               &                               & ${\bf (1,\bar{3},2)}$         & ${\bf (1,\bar{3},2)}$         \\
${\bf (\bar{3},1)}$             & ${\bf (1,\bar{3},1)}$         & ${\bf (1,\bar{3},1)}$         & ${\bf (1,\bar{3},1)}$         &                               & ${\bf (1,\bar{3},2)}$         \\
${\bf (\bar{3},1)}$             & ${\bf (1,\bar{3},1)}$         & ${\bf (1,\bar{3},1)}$         &                               &                               &                               \\
				& ${\bf (1,\bar{3},1)}$         &                               &                               &                               & ${\bf (1,3,1)}$               \\
\hline
${\bf (1,2)}$                   &                               & ${\bf (1,1,2)}$               &                               & ${\bf (1,1,2)}$               &                               \\
${\bf (1,1)}$                   & ${\bf (1,1,1)}$               & ${\bf (1,1,1)}$               & ${\bf (1,1,1)}$               & ${\bf (1,1,1)}$               & ${\bf (1,1,1)}$               \\
\hline\hline
$\# Q=1(\times 3)$      & $\# Q=3$ & $\# Q=4-1$ & $\# Q=5-2$ & $\# Q=6-3$ & $\# Q=7-4$\\
$\# \bar{q}=2(\times 3)$ & $\# \bar{q}=3+3$& $\# \bar{q}=4+2$ & $\# \bar{q}=5+1$ & $\# \bar{q}=6$ & $\# \bar{q}=7-1$ \\
$\# L=1(\times 3)$      & $\# L=3$ & $\# L=4\pm 1$ & $\# L=5$ & $\# L=6\pm 1$ & $\# L=7$ \\
\hline\hline
\end{tabular}

\bigskip
\bigskip

\newpage

Table 2. Embedding the three family SM.

\vspace{.1in}

\noindent
\begin{tabular}{|c|c|cc|cc|}
\hline\hline
$SU(4)_A\otimes SU(3)_C\otimes SU(2)_L$ Rep. & $U(1)_X$ &
\multicolumn{2}{|c|}{$U(1)_Y$ states} & \multicolumn{2}{|c|}{3-family $U(1)_Y$ states}\\
\hline
${\bf (4,3,2)}$         &       $1$  &  $\alpha - \beta \pm \gamma$     & $\alpha + \beta \pm \delta$   & (3)\  $\alpha - \beta$     & $\alpha + 3\beta$  \\
${\bf (1,\bar{3},2)}$   &       $a$  &  \multicolumn{2}{|c|}{$a\alpha$}                    &                       & $-\alpha - 3\beta$ \\ \hline
${\bf (\bar{4},\bar{3},1)}$ &   $x$  &  $x\alpha + \beta \pm \gamma$    & $x\alpha - \beta \pm \delta$  & (3)\ $x\alpha + \beta$     & $x\alpha - 3\beta$ \\
${\bf (1,\bar{3},1)}$   &       $b$ &   \multicolumn{2}{|c|}{$b\alpha$}                    &                       & $x\alpha - 3\beta$ \\
${\bf (1,\bar{3},1)}$   &       $c$ &   \multicolumn{2}{|c|}{$c\alpha$}                    &                       & $x\alpha - 3\beta$ \\ \hline
${\bf (\bar{4},1,2)}$   &       $y$  &  $y\alpha + \beta \pm \gamma$    & $y\alpha - \beta \pm \delta$  & (3)\ $y\alpha + \beta$     & $y\alpha - 3\beta$ \\
${\bf (1,1,2)}$         &       $k$ &   \multicolumn{2}{|c|}{$k\alpha$}                    &                       & $-y\alpha + 3\beta$        \\  \hline
${\bf (\bar{4},1,1)}$   &       $z$  &  $z\alpha + \beta \pm \gamma$    & $z\alpha - \beta \pm \delta$  & (3)\ $z\alpha + \beta$     & $z\alpha - 3\beta$ \\
${\bf (1,1,1)}$         &       $s$ &   \multicolumn{2}{|c|}{$s\alpha$}                    &                       & $-z\alpha + 3\beta$        \\
\hline\hline
\end{tabular}



\bigskip
\bigskip

Table 3. {\it Model IA} -- representation structures, anomaly cancellations
and the hypercharge embedding.

\vspace{.1in}

\scriptsize

\noindent
\begin{tabular}{|c|c|r|r|r|r|r|cc|}
\hline\hline
$SU(4)_A\otimes SU(3)_C\otimes SU(2)_L$ Rep. & $U(1)_X$ &
\multicolumn{5}{|c|}{Gauge anomalies} &  \multicolumn{2}{|c|}{$U(1)_Y$
states}   \\ \hline
&  &    $U(1)$-grav. & $[SU(4)]^2U(1)$  & $[SU(3)]^2U(1)$ & $[SU(2)]^2U(1)$ &  $[U(1)]^3$ &                            & \\ \hline
${\bf (4,3,2)}$         &       {\bf 5}  &     120  &   30  &   40 &    60 &    3000    &       3\ {\bf 1}($Q$)         & {\bf 7}($Q^{'}$) \\
${\bf (\bar{4},\bar{3},1)}$ &   {\bf 1}  &      12  &   3 &     4 &      &      12      &       3\ {\bf 2}($\bar{d}$)   & {\bf -4}($\bar{u}$) \\
${\bf (\bar{4},1,2)}$   &       {\bf -9}  &     -72  &  -18  &  &       -36 &   -5832   &       3\ {\bf -3}($L$)        & {\bf -9}($L^{'}$) \\
${\bf (\bar{4},1,1)}$   &       {\bf 9}  &      36  &   9  &    &        &      2916    &       3\ {\bf 6}($E$)       & {\bf 0}($N$) \\  \hline
${\bf (1,\bar{3},2)}$   &       {\bf -14} &     -84 &   &       -28 &   -42 &   -16464  &       \multicolumn{2}{|c|}{{\bf -7}($\bar{Q}^{'}$)} \\
${\bf (1,\bar{3},1)}$   &       {\bf -8} &      -24 &   &       -8  &    &      -1536   &       \multicolumn{2}{|c|}{{\bf -4}($\bar{u}$)} \\
${\bf (1,\bar{3},1)}$   &       {\bf -8} &      -24 &   &       -8  &    &      -1536   &       \multicolumn{2}{|c|}{{\bf -4}($\bar{u}$)} \\ \hline
${\bf (1,1,2)}$         &       {\bf 18}  &     36  &   &       &       18  &   11664   &       \multicolumn{2}{|c|}{{\bf 9}($\bar{L}^{'}$)} \\
${\bf (1,1,1)}$         &       {\bf 0} &           &   &       &       &               &       &       \\  \hline
\multicolumn{2}{|r|}{\it subtotal}      &       0   &   24  &   0  &    0  &    -7776   &                               & \\            \hline
${\bf (6,1,1)}$         &       {\bf -12} &     -72 &   -24 &   &        &      -10368  &       3\ {\bf -3}($S$)        & 3\ {\bf -9}($S^{'}$) \\
$3\ {\bf (1,1,1)}$      &       {\bf 6} &       18  &   &       &       &       648     &       \multicolumn{2}{|c|}{3\ {\bf 3}($\bar{S}$)} \\
$3\ {\bf (1,1,1)}$      &       {\bf 18} &      54 &    &       &       &       17496   &       \multicolumn{2}{|c|}{3\ {\bf 9}($\bar{S}^{'}$)} \\      \hline
\multicolumn{2}{|r|}{\it Total}         &       0   &   0  &    0  &    0  &    0       &                               & \\
\hline\hline
\end{tabular}

\clearpage


Table 4. Explicit contents of four $SU(4)_A\otimes SU(3)_C\otimes SU(2)_L\otimes U(1)_X$
models.

\vspace{.3in}


\noindent
\begin{tabular}{||c|cc||c|cc||c|cc||c|cc||}
\hline\hline
\multicolumn{3}{||c||}{{\it Model IA}}     & 
\multicolumn{3}{|c||}{{\it Model IIA}} & 
\multicolumn{3}{|c||}{{\it Model IIB}} &
\multicolumn{3}{|c||}{{\it Model Im}}  \\   \hline
${\bf (4,3,2,5)}$               &       3\ {\bf 1}($Q$)         & {\bf 7}($Q^{'}$)              & 
${\bf (4,3,2,1)}$             & 3\ {\bf 1}($Q$)        & {\bf -5}($Q^{''}$)           & 
${\bf (4,3,2,1)}$             & 3\ {\bf 1}($Q$)          & {\bf -5}($Q^{''}$)        &
${\bf (4,3,2,5)}$               &       3\ {\bf 1}($Q$)         & {\bf 7}($Q^{'}$)  \\
${\bf (\bar{4},\bar{3},1,1)}$   &       3\ {\bf 2}($\bar{d}$)   & {\bf -4}($\bar{u}$)           & ${\bf (\bar{4},\bar{3},1,5)}$ & 3\ {\bf -4}($\bar{u}$) & {\bf 2}($\bar{d}$)           & 
${\bf (\bar{4},\bar{3},1,5)}$ & 3\ {\bf -4}($\bar{u}$)   & {\bf 2}($\bar{d}$)      & 
${\bf (\bar{4},\bar{3},1,1)}$   &       3\ {\bf 2}($\bar{d}$)   & {\bf -4}($\bar{u}$)    \\
${\bf (\bar{4},1,2,-9)}$        &       3\ {\bf -3}($L$)        & {\bf -9}($L^{'}$)             & ${\bf (\bar{4},1,2,3)}$       & 3\ {\bf -3}($L$)   & {\bf 3}($\bar{L}$)               & 
${\bf (\bar{4},1,2,3)}$       & 3\ {\bf -3}($L$)     & {\bf 3}($\bar{L}$)      &     
${\bf (\bar{4},1,2,-9)}$        &       3\ {\bf -3}($L$)        & {\bf -9}($L^{'}$)    \\
${\bf (\bar{4},1,1,9)}$         &       3\ {\bf 6}($E$)       & {\bf 0}($N$)                  & ${\bf (\bar{4},1,1,-15)}$     & 3\ {\bf 6}($E$)      & {\bf 12}($S^{''}$)           	& 
${\bf (\bar{4},1,1,9)}$       & 3\ {\bf -6}($\bar{E}$) & {\bf 0}($N$)       &       
 ${\bf (\bar{4},1,1,-15)}$   & 3\ {\bf -6}($\bar{E}$)      & {\bf -12}($\bar{S}^{''}$)    \\
${\bf (6,1,1,-12)}$             &       3\ {\bf -3}($S$)        & 3\ {\bf -9}($S^{'}$)          & ${\bf (6,1,1,-6)}$            & 3\ {\bf 6}($E$)      & 3\ {\bf 0}($N$)              & 
${\bf (6,1,1,-18)}$           & 3\ {\bf 6}($E$)        & 3\ {\bf 12}($S^{''}$)     &
  &   &   \\
\hline
${\bf (1,\bar{3},2,-14)}$       &      
\multicolumn{2}{|c||}{{\bf -7}($\bar{Q}^{'}$)}           & ${\bf (1,\bar{3},2,-10)}$     & 
\multicolumn{2}{|c||}{{\bf 5}($\bar{Q}^{''}$)}         & ${\bf (1,\bar{3},2,-10)}$     & 
\multicolumn{2}{|c||}{{\bf 5}($\bar{Q}^{''}$)}        &
${\bf (1,\bar{3},2,-14)}$       &      
\multicolumn{2}{|c||}{{\bf -7}($\bar{Q}^{'}$)}      \\
${\bf (1,\bar{3},1,-8)}$        &       \multicolumn{2}{|c||}{{\bf -4}($\bar{u}$)}               & ${\bf (1,\bar{3},1,-4)}$      & 
\multicolumn{2}{|c||}{{\bf 2}($\bar{d}$)}              & ${\bf (1,\bar{3},1,-4)}$      & 
\multicolumn{2}{|c||}{{\bf 2}($\bar{d}$)}             &
${\bf (1,\bar{3},1,-8)}$        &       \multicolumn{2}{|c||}{{\bf -4}($\bar{u}$)}  \\
${\bf (1,\bar{3},1,-8)}$        &       
\multicolumn{2}{|c||}{{\bf -4}($\bar{u}$)}               & ${\bf (1,\bar{3},1,-4)}$      & 
\multicolumn{2}{|c||}{{\bf 2}($\bar{d}$)}              & ${\bf (1,\bar{3},1,-4)}$      & 
\multicolumn{2}{|c||}{{\bf 2}($\bar{d}$)}             &
${\bf (1,\bar{3},1,-8)}$        &       \multicolumn{2}{|c||}{{\bf -4}($\bar{u}$)} \\ \hline
${\bf (1,1,2,18)}$              &       
\multicolumn{2}{|c||}{{\bf 9}($\bar{L}^{'}$)}            & ${\bf (1,1,2,6)}$             & 
\multicolumn{2}{|c||}{{\bf -3}($L$)}               &  ${\bf (1,1,2,6)}$
            &  \multicolumn{2}{|c||}{{\bf -3}($L$)}            &
${\bf (1,1,2,18)}$              &       
\multicolumn{2}{|c||}{{\bf 9}($\bar{L}^{'}$)}    \\
$3\ {\bf (1,1,1,6)}$            &       
\multicolumn{2}{|c||}{3\ {\bf 3}($\bar{S}$)}             & ${\bf (1,1,1,24)}$            & 
\multicolumn{2}{|c||}{{\bf -12}($\bar{S}^{''}$)}      	&  $3\ {\bf (1,1,1,24)}$         &
\multicolumn{2}{|c||}{3\ {\bf -12}($\bar{S}^{''}$)}  &
${\bf (1,1,1,24)}$            & 
\multicolumn{2}{|c||}{{\bf 12}($S^{''}$)}      	  \\
$3\ {\bf (1,1,1,18)}$           &       
\multicolumn{2}{|c||}{3\ {\bf 9}($\bar{S}^{'}$)}         & $3\ {\bf (1,1,1,12)}$         & 
\multicolumn{2}{|c||}{3\ {\bf -6}($\bar{E}$)}        &  $3\ {\bf (1,1,1,-12)}$        & 
\multicolumn{2}{|c||}{3\ {\bf 6}($E$)}           &
$6\ {\bf (1,1,1,12)}$        & 
\multicolumn{2}{|c||}{6\ {\bf 6}($E$)}    \\
\hline\hline
\end{tabular}

\bigskip
\bigskip


Table 5. Minimal  $SU(5)_A\otimes SU(3)_C\otimes SU(2)_L\otimes U(1)_X$
models.

\vspace{.3in}


\noindent
\begin{tabular}{|c|r|cc|r|cc|r|cc|r|cc|}
\hline\hline
 				&	\multicolumn{3}{|c|}{{\it Scheme IA }}   		& \multicolumn{3}{|c|}{{\it Scheme IB}}   			& 
\multicolumn{3}{|c|}{{\it Scheme IIA }}   			& \multicolumn{3}{|c|}{{\it Scheme IIB}}  	\\
\cline{2-13}
 				& $U(1)_X$	& \multicolumn{2}{|c|}{$Y$-states}		& $U(1)_X$	&
\multicolumn{2}{|c|}{$Y$-states} 		& $U(1)_X$	& \multicolumn{2}{|c|}{$Y$-states}		& $U(1)_X$	& \multicolumn{2}{|c|}{$Y$-states} 	  \\   \hline
${\bf (5,3,2)}$                 & 17    & 3\ {\bf 1}($Q$)       & 2\ {\bf 7}($Q^{'}$)           & 
17    & 3\ {\bf 1}($Q$)       & 2\ {\bf 7}($Q^{'}$)          	& 7     & 3\ {\bf 1}($Q$)        & 2\ {\bf -5}($Q^{''}$)        & 7     & 3\ {\bf 1}($Q$)    & 2\ {\bf -5}($Q^{''}$)  \\
${\bf (\bar{5},\bar{3},1)}$     & -2    & 3\ {\bf 2}($\bar{d}$) & 2\ {\bf -4}($\bar{u}$)        & 
-2    & 3\ {\bf 2}($\bar{d}$) & 2\ {\bf -4}($\bar{u}$)        & 8     & 3\ {\bf -4}($\bar{u}$) & 2\ {\bf 2}($\bar{d}$)        & 8     & 3\ {\bf -4}($\bar{u}$)   & 2\ {\bf 2}($\bar{d}$)  \\
${\bf (\bar{5},1,2,)}$          & -27   & 3\ {\bf -3}($L$)      & 2\ {\bf -9}($L^{'}$)          & 3     & 3\ {\bf 3}($\bar{L}$) & 2\ {\bf -3}($L$)             	& 
3     & 3\ {\bf -3}($L$)   	 & 2\ {\bf 3}($\bar{L}$)        & 33    & 3\ {\bf -9}($L{'}$)  & 2\ {\bf -3}($L$)       \\
${\bf (\bar{5},1,1)}$           & -42   & 3\ {\bf -6}($\bar{E}$) & 2\ {\bf -12}($\bar{S}^{''}$) & -102  & 3\ {\bf -18}($\bar{T}$) & 2\ {\bf -24} ($\bar{T}^{'}$) &
-72   & 3\ {\bf 12}($S^{''}$)    & 2\ {\bf 18}($T$)        & -132  & 3\ {\bf 24} ($T{'}$)   & 2\ {\bf 30}($T^{''}$)  \\
\hline
$2\ {\bf (1,\bar{3},2)}$        & -35   & \multicolumn{2}{|c|}{2\ {\bf -7}($\bar{Q}^{'}$)}      & -35   & \multicolumn{2}{|c|}{2\ {\bf -7}($\bar{Q}^{'}$)}      & 
-25   & \multicolumn{2}{|c|}{2\ {\bf 5}($\bar{Q}^{''}$)}      & -25   & \multicolumn{2}{|c|}{2\ {\bf 5}($\bar{Q}^{''}$)}  \\
${\bf (1,\bar{3},1)}$           & -20   & \multicolumn{2}{|c|}{{\bf -4}($\bar{u}$)}             & -20   & \multicolumn{2}{|c|}{{\bf -4}($\bar{u}$)}             & 
-10   & \multicolumn{2}{|c|}{{\bf 2}($\bar{d}$)}              & -10   & \multicolumn{2}{|c|}{{\bf 2}($\bar{d}$)}          \\
\hline
$2\ {\bf (1,1,2)}$              & 45    & \multicolumn{2}{|c|}{2\ {\bf 9}($\bar{L}^{'}$)}       & -15      & \multicolumn{2}{|c|}{2\ {\bf -3}($L$)}      	& 
15    & \multicolumn{2}{|c|}{{\bf -3}($L$)}               	& -45   & \multicolumn{2}{|c|}{2\ {\bf 9}($\bar{L}^{'}$)}    \\
${\bf (1,1,2)}$                 &     	& &          						& -15      &
\multicolumn{2}{|c|}{{\bf -3}($L$)}      		&     	& &                					&
-45   & \multicolumn{2}{|c|}{{\bf 9}($\bar{L}^{'}$)}  \\
${\bf (1,1,2)}$                 &     	& &          						& -15      &
\multicolumn{2}{|c|}{{\bf -3}($L$)}      		&     	& &                					&
15    & \multicolumn{2}{|c|}{{\bf -3}($L$)}         	\\ \hline
$3\ {\bf (1,1,1)}$              & 30    & \multicolumn{2}{|c|}{3\ {\bf 6}($E$)}           	& 30      & \multicolumn{2}{|c|}{3\ {\bf 6}($E$)}           	& 
-30   & \multicolumn{2}{|c|}{3\ {\bf 6}($E$)}      		& -30   & 
\multicolumn{2}{|c|}{3\ {\bf 6}($E$)}       \\
$3\ {\bf (1,1,1)}$              & 30    & \multicolumn{2}{|c|}{3\ {\bf 6}($E$)}       	& 90      & \multicolumn{2}{|c|}{3\ {\bf 18}($T$)}      	&  
60   & \multicolumn{2}{|c|}{3\ {\bf -12}($\bar{S}^{''}$)}    & 120   & \multicolumn{2}{|c|}{3\ {\bf -24}($\bar{T}^{'}$)}       \\
$2\ {\bf (1,1,1)}$              & 60    & \multicolumn{2}{|c|}{2\ {\bf12}($S^{''}$)}       	& 
120     & \multicolumn{2}{|c|}{2\ {\bf 24}($T^{'}$)}          &  90   &
\multicolumn{2}{|c|}{2\ {\bf -18}($\bar{T}$)}        	& 150   &
\multicolumn{2}{|c|}{2\ {\bf -30}($\bar{T}^{''}$)}      \\ \hline\hline
\end{tabular}

\newpage

\bigskip
\bigskip


Table 6. Minimal  $SU(6)_A\otimes SU(3)_C\otimes SU(2)_L\otimes U(1)_X$
models.

\vspace{.3in}


\noindent
\begin{tabular}{|c|r|cc|r|cc|r|cc|r|cc|}
\hline\hline
 				&	\multicolumn{3}{|c|}{{\it Scheme IA }}   		& \multicolumn{3}{|c|}{{\it Scheme IB}}   			& 
\multicolumn{3}{|c|}{{\it Scheme IIA }}   			& \multicolumn{3}{|c|}{{\it Scheme IIB}}  	\\
\cline{2-13}
 				& $U(1)_X$	& \multicolumn{2}{|c|}{$Y$-states}		& $U(1)_X$
& \multicolumn{2}{|c|}{$Y$-states} 		& 
$U(1)_X$	& \multicolumn{2}{|c|}{$Y$-states}		& $U(1)_X$	& \multicolumn{2}{|c|}{$Y$-states} 	  \\   \hline
${\bf (6,3,2)}$                 & -4    & 3\ {\bf 1}($Q$)       & 3\ {\bf 7}($Q^{'}$)           & -4    & 3\ {\bf 1}($Q$)       & 3\ {\bf 7}($Q^{'}$)          	& 
2     & 3\ {\bf 1}($Q$)        & 3\ {\bf -5}($Q^{''}$)        & 2     & 3\ {\bf 1}($Q$)    & 3\ {\bf -5}($Q^{''}$)  \\
${\bf (\bar{6},\bar{3},1)}$     & 1    	& 3\ {\bf 2}($\bar{d}$) & 3\ {\bf -4}($\bar{u}$)        & 1    	& 3\ {\bf 2}($\bar{d}$) & 3\ {\bf -4}($\bar{u}$)        & 
1     & 3\ {\bf -4}($\bar{u}$) & 3\ {\bf 2}($\bar{d}$)        & 1     & 3\ {\bf -4}($\bar{u}$)   & 3\ {\bf 2}($\bar{d}$)  \\
${\bf (\bar{6},1,2,)}$          & 0   	& 3\ {\bf 3}($\bar{L}$) & 3\ {\bf -3}($L$)          	& 6     & 3\ {\bf -3}($L$) 	& 3\ {\bf -9}($L^{'}$)          & 
6     & 3\ {\bf -9}($L^{'}$)   & 3\ {\bf -3}($L$)        	& 0    	& 3\ {\bf -3}($L$)         & 3\ {\bf 3}($\bar{L}$)       \\
${\bf (\bar{6},1,1)}$           & 21   	& 3\ {\bf -18}($\bar{T}$)  & 3\ {\bf -24}($\bar{T}^{'}$)  & 
9  	& 3\ {\bf -6}($\bar{E}$)  & 3\ {\bf -12}($\bar{S}^{''}$) & 
-27   & 3\ {\bf 24}($T^{'}$) & 3\ {\bf 30}($T^{''}$)       & -15  	& 3\ {\bf 12}($S^{''}$)    & 3\ {\bf 18}($T$)            \\
\hline
$3\ {\bf (1,\bar{3},2)}$        & 7   	& \multicolumn{2}{|c|}{3\ {\bf -7}($\bar{Q}^{'}$)}      & 7   	& \multicolumn{2}{|c|}{3\ {\bf -7}($\bar{Q}^{'}$)}      & 
-5    & \multicolumn{2}{|c|}{3\ {\bf 5}($\bar{Q}^{''}$)}      & -5   	& \multicolumn{2}{|c|}{3\ {\bf 5}($\bar{Q}^{''}$)}  \\
$3\ {\bf (1,1,2)}$              & 3    	& \multicolumn{2}{|c|}{3\ {\bf 3}($L$)}       		& -9    & \multicolumn{2}{|c|}{3\ {\bf 9}($\bar{L}^{'}$)}      	& 
-9    & \multicolumn{2}{|c|}{3\ {\bf 9}($\bar{L}^{'}$)}       & 3   	& \multicolumn{2}{|c|}{3\ {\bf -3}($L$)}    \\
\hline
$3\ {\bf (1,1,1)}$              & -6    & \multicolumn{2}{|c|}{3\ {\bf 6}($E$)}           	& -6      & \multicolumn{2}{|c|}{3\ {\bf 6}($E$)}           	& 
-6    & \multicolumn{2}{|c|}{3\ {\bf 6}($E$)}      		& -6   & \multicolumn{2}{|c|}{3\ {\bf 6}($E$)}       \\
$3\ {\bf (1,1,1)}$              & -18   & \multicolumn{2}{|c|}{3\ {\bf 18}($T$)}       & 
-6      & \multicolumn{2}{|c|}{3\ {\bf 6}($E$)}      	&  
24   & \multicolumn{2}{|c|}{3\ {\bf -24}($\bar{T}^{'}$)}     & 12   &
\multicolumn{2}{|c|}{3\ {\bf -12}($\bar{S}^{''}$)}    \\
$3\ {\bf (1,1,1)}$              & -24   & \multicolumn{2}{|c|}{3\ {\bf 24}($T{'}$)}       	& 
-12     & \multicolumn{2}{|c|}{3\ {\bf 12}($S^{''}$)}         &  30   &
\multicolumn{2}{|c|}{3\ {\bf -30}($\bar{T}^{''}$)}    & 18   &
\multicolumn{2}{|c|}{3\ {\bf -18}($\bar{T}$)}            \\ \hline\hline
\end{tabular}


\bigskip
\bigskip

Table 7. Minimal  $SU(3)_H\otimes SU(3)_C\otimes SU(2)_L\otimes U(1)_Y$
models.

\vspace{.3in}

\footnotesize

\noindent
\begin{tabular}{|c|c|c|}
\hline\hline
				& {\it Scheme I }   & {\it Scheme II}   \\
				& $U(1)_Y$-states		& $U(1)_Y$-states	\\  \hline
${\bf (3,3,2)}$                 & 3\ {\bf 1}($Q$)               & 3\ {\bf 1}($Q$)       	\\
${\bf (\bar{3},\bar{3},1)}$     & 3\ {\bf 2}($\bar{d}$) 	& 3\ {\bf -4}($\bar{u}$)        \\
${\bf (\bar{3},1,2,)}$          & 3\ {\bf -3}($L$)          	& 3\ {\bf -3}($L$) 		\\
${\bf (\bar{3},1,1)}$           & 3\ {\bf -6}($\bar{E}$)     	& 3\ {\bf -12}($\bar{S}^{''}$)  \\
\hline
$3\ {\bf (1,\bar{3},1)}$        & 3\ {\bf -4}($\bar{u}$) 	& 3\ {\bf 2}($\bar{d}$)     	\\
$3\ {\bf (1,1,1)}$              & 3\ {\bf 6}($E$)           	& 3\ {\bf 6}($E$)           	\\
$3\ {\bf (1,1,1)}$              & 3\ {\bf 6}($E$)        	& 3\ {\bf 12}($S^{''}$)       	\\
\hline\hline
\end{tabular}

\clearpage

\bigskip

\scriptsize
 
Table 8a. A possible pattern of mass generation, starting with  $\Phi = {\bf (15,1,2,-6)}$.

\vspace{.3in}
\noindent
\begin{tabular}{|c||cc|cc|cc|}\hline\hline
  &  \multicolumn{6}{|c|}{Effective EW-doublet VEVs and mass terms generated} \\  \cline{2-7} 
Extra singlet VEV(s)  &   $\left \langle \Phi _{eff}  \right \rangle$ 
& $SU(4)_A\otimes U(1)_X$-rep.  &  VEV(s)  & Mass term(s)  &  VEV(s)  & Mass term(s) \\
\cline{2-7} 
  & $\left \langle \Phi  \right \rangle$  & $15_{-6}$ & ${\bf \left \langle (1,3_0)_{0,-6}\right \rangle }$
& $M^{(3)}_{33}$ ,  $M^{(4)}_{34}$  &  ${\bf \left \langle (1,3_-)_{0,-6}\right \rangle }$ 
& $M^{(4)}_{33}$  \\  \hline
$\left \langle \phi_0  \right \rangle --- {\bf \left \langle (1,2_+)_{-1,9}\right \rangle }$
& $\left \langle \Phi  \right \rangle v_0 / M_0 $ 
&  $15_{-6}\times \bar{4}_9 = \bar{4}_3$ &   ${\bf \left \langle (1,2_+)_{-1,3}\right \rangle }$
&  $M^{(4)}_{14}, M^{(4)}_{24}$  &  ${\bf \left \langle (1,2_-)_{-1,3}\right \rangle }$ 
&  $M^{(4)}_{13}, M^{(4)}_{23}$  \\
$\left \langle \phi_a  \right \rangle --- {\bf \left \langle (2_{\pm},1)_{1,-3}\right \rangle }$
& $\left \langle \Phi ^{\dag}  \right \rangle v_0^{*} v_a / M_0^2$  
&  $4_{-3}\times \bar{4}_{-3} = 15_{-6}$  
&   ${\bf \left \langle (2_{\pm},2_-)_{2,-6}\right \rangle }$
&  $M^{(4)}_{31},  M^{(4)}_{32}$   &  ${\bf \left \langle (2_{\pm},2_+)_{2,-6}\right \rangle }$ 
&  $M^{(3)}_{31},  M^{(3)}_{32}$  \\  \hline\hline
\end{tabular}

\bigskip
\bigskip

Table 8b.  An alternative possible pattern of mass generation,  starting with  $\Phi = {\bf (\bar{4},1,2,3)}$.

\vspace{.3in}
\noindent
\begin{tabular}{|c||cc|cc|cc|}
\hline\hline
  &  \multicolumn{6}{|c|}{Effective EW-doublet VEVs and mass terms generated} \\  \cline{2-7} 
Extra singlet VEV(s)  &   $\left \langle \Phi _{eff}  \right \rangle$ 
& $SU(4)_A\otimes U(1)_X$-rep.  &  VEV(s)  & Mass term(s)  &  VEV(s)  & Mass term(s) \\
\cline{2-7} 
\cline{2-7} 
 & $\left \langle \Phi  \right \rangle$  &  $\bar{4}_3$
& ${\bf \left \langle (1,2_-)_{-1,3}\right \rangle }$
& $M^{(4)}_{13} ,  M^{(4)}_{14}$  &  ${\bf \left \langle (1,2_+)_{-1,3}\right \rangle }$
& $M^{(4)}_{23} ,  M^{(4)}_{24}$  \\ \hline
$\left \langle \phi_0^{\dag}  \right \rangle --- {\bf \left \langle ((1,2_+)_{-1,9})^{\dag}\right \rangle }$
& $\left \langle \Phi  \right \rangle v_0^* / M_0 $ 
& $\bar{4}_3 \times 4_{-9} = 15_{-6}$   &   ${\bf \left \langle (1,3_-)_{0,-6}\right \rangle }$ 
&  $M^{(4)}_{33}$  &   ${\bf \left \langle (1,3_0)_{0,-6}\right \rangle }$ 
&  $M^{(3)}_{33}, M^{(4)}_{34}$  \\
$\left \langle \phi_a  \right \rangle --- {\bf \left \langle (2_{\pm},1)_{1,-3}\right \rangle }$
& $\left \langle \Phi ^{\dag}  \right \rangle  v_a / M_0$  
&  $4_{-3}\times \bar{4}_{-3} = 15_{-6}$  
&   ${\bf \left \langle (2_{\pm},2_+)_{2,-6}\right \rangle }$ 
&  $M^{(3)}_{31},  M^{(3)}_{32}$   &  ${\bf \left \langle (2_{\pm},2_-)_{2,-6}\right \rangle }$
&  $M^{(4)}_{31},  M^{(4)}_{32}$    \\ \hline\hline
\end{tabular}

\clearpage

\begin{flushleft}
\begin{figure}[h]


\setlength{\unitlength}{1.0cm}

\begin{picture}(15,13)

\thicklines

\put(5.08,8){\vector(1,0){2}}
\put(7.08,8){\line(1,0){0.84}}
\multiput(8,8)(0,0.3){11}{\line(0,1){0.25}}
\put(7.85,11.2){${\bf \times}$}
\put(8.08,8){\line(1,0){0.84}}
\put(10.22,8){\vector(-1,0){1.3}}
\put(2,7.9){${\bf 4_5}$}
\put(7.5,11.6){$ \left \langle \Phi = {\bf 15_{-6}} \right \rangle $}
\put(10.5,7.9){${\bf 1_{-8}}$}
\put(2.78,8){\vector(1,0){1.3}}
\put(4.08,8){\line(1,0){0.84}}
\multiput(5,8)(0,0.3){11}{\line(0,1){0.25}}
\put(4.85,11.2){${\bf \times}$}
\put(4.5,11.6){$ \left \langle \phi_0 = {\bf \bar{4}_{9}} \right \rangle $}
\put(6.5,7.5){${\bf 15_{14}}$}

\put(1.5, 6){{\it  OR}}

\put(2.78,2){\vector(1,0){1.3}}
\put(4.08,2){\line(1,0){0.84}}
\multiput(5,2)(0,0.3){11}{\line(0,1){0.25}}
\put(4.85,5.2){${\bf \times}$}
\put(5.08,2){\line(1,0){0.84}}
\put(7.92,2){\vector(-1,0){2}}
\multiput(8,2)(0,0.3){11}{\line(0,1){0.25}}
\put(7.85,5.2){${\bf \times}$}
\put(8.08,2){\line(1,0){0.84}}
\put(10.22,2){\vector(-1,0){1.3}}
\put(2,1.9){${\bf 4_5}$}
\put(4.5,5.6){$ \left \langle  \Phi = {\bf 15_{-6}} \right \rangle $}
\put(6.2,1.5){$ {\bf \bar{4}_{1}}$}
\put(7.5,5.6){$ \left \langle \phi_0 = {\bf \bar{4}_{9}} \right \rangle $}
\put(10.5,1.9){${\bf 1_{-8}}$}

\end{picture}

\caption{Illustrations of Froggatt-Nielsen mechanism for quark mass generation. }

\label{Fig. 1}

\end{figure}
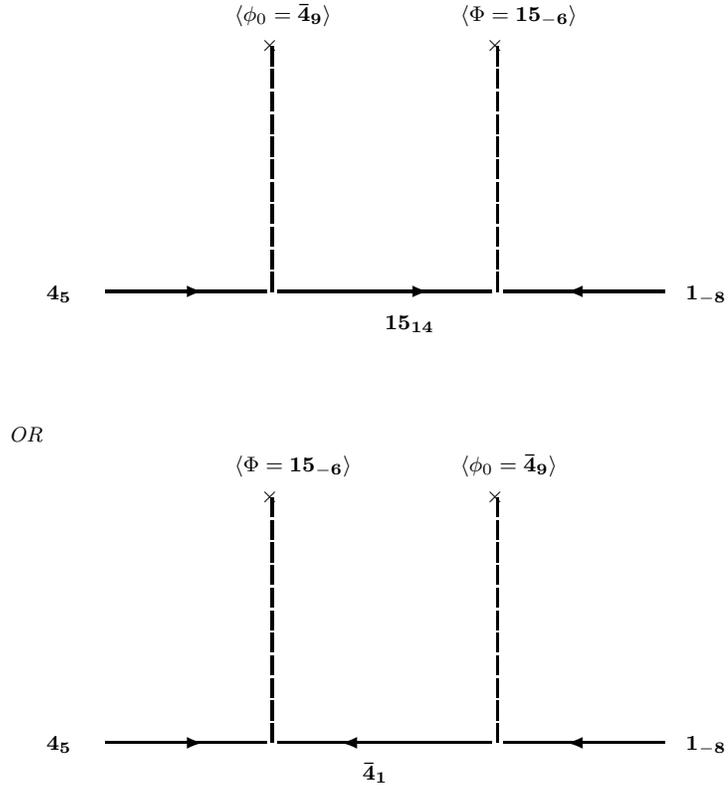

\end{flushleft}

\clearpage

\begin{flushleft}
\begin{figure}[h]


\setlength{\unitlength}{1.0cm}

\begin{picture}(8,11)

\thicklines

\LARGE

\multiput(3.88,7.8)(0.385,0){7}{$\sim$}
\put(2,10){\line(1,-1){1.2}}
\put(4,8){\line(-1,-1){1.2}}
\put(4,8){\vector(-1,1){1}}
\put(2,6){\vector(1,1){1}}

\multiput(3.88,1.8)(0.385,0){7}{$\sim$}
\put(2,4){\line(1,-1){1.2}}
\put(4,2){\line(-1,-1){1.2}}
\put(4,2){\vector(-1,1){1}}
\put(2,0){\vector(1,1){1}}



\normalsize

\put(6.2,8.2){ ${\bf (2_{\mp},2_-)_2}$}
\put(0,6){${\bf (1,2_+)_{-1}}$}
\put(0,10){${\bf (2_{\pm},1)_{1}}$}

\put(6.2,2.2){${\bf (1,3_+)_0}$}
\put(0,0){${\bf (1,2_+)_{-1}}$}
\put(0,4){${\bf (1,2_-)_{-1}}$}

\end{picture}

\caption{Charged gauge boson vertices with quark singlets. }

\label{Fig. 2}

\end{figure}
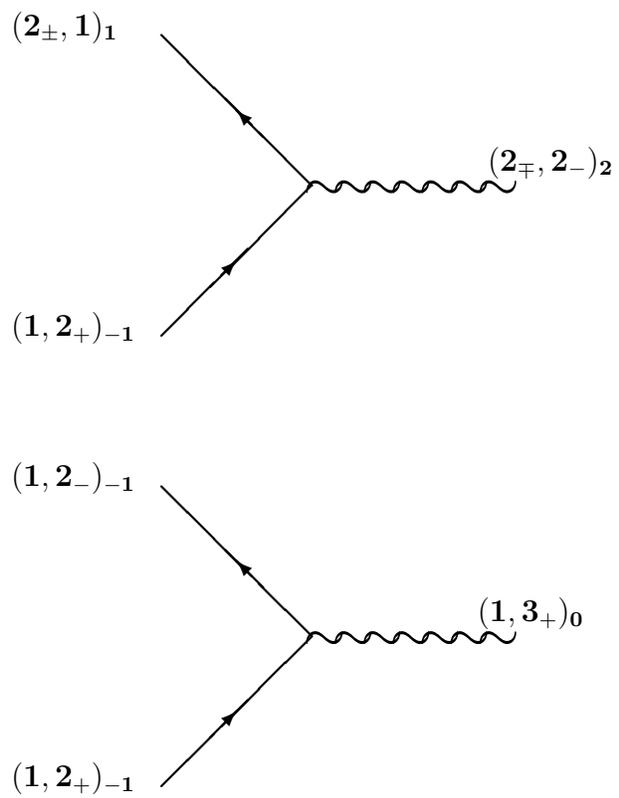

\end{flushleft}

\clearpage

\begin{flushleft}
\begin{figure}[h]


\setlength{\unitlength}{1.0cm}

\begin{picture}(11,7)

\thicklines

\LARGE

\multiput(3.88,3.8)(0.385,0){8}{$\sim$}
\put(2,6){\vector(1,-1){1}}
\put(4,4){\vector(-1,-1){1}}
\put(4,4){\line(-1,1){1.2}}
\put(2,2){\line(1,1){1.2}}

\put(7.05,4){\vector(1,1){1}}
\put(7.05,4){\line(1,-1){1.2}}
\put(9.05,2){\vector(-1,1){1}}
\put(9.05,6){\line(-1,-1){1.2}}

\normalsize
\put(2.3,1.8){${\bf E(\sim {\it l}^+)}$}
\put(1.5,6.1){${\bf N}$}

\put(9.2,5.8){${\bf \bar{u}_i}$}
\put(9.2,1.8){${\bf \bar{d}_j}$}

\end{picture}

\caption{Characteristic decay mode of a right-handed neutrino. }

\label{Fig. 3}

\end{figure}
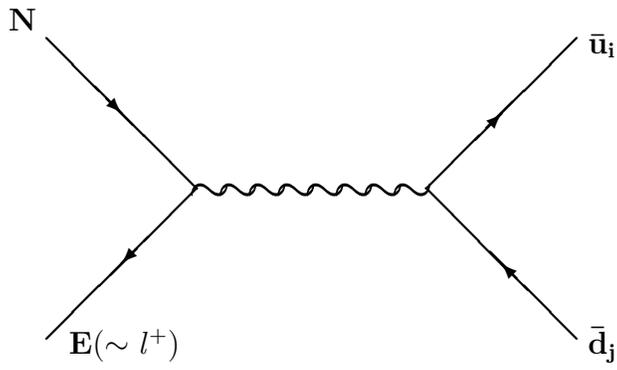

\end{flushleft}

\end{document}